\documentclass[journal]{IEEEtran}

\usepackage{amsmath,amsfonts}
\usepackage{amssymb}
\usepackage{array}
\usepackage{graphicx}
\usepackage{booktabs}
\usepackage{multirow}
\usepackage{tabularx}
\usepackage{longtable}
\usepackage{url}
\usepackage{cite}
\usepackage[ruled,vlined,linesnumbered]{algorithm2e}
\usepackage[hidelinks]{hyperref}
\SetKwInOut{Input}{Input}
\SetKwInOut{Output}{Output}
\DontPrintSemicolon
\SetKwFor{ForEach}{\textbf{for} each}{do}{end}
\usepackage{xcolor}
\usepackage{microtype}

\begin{document}
	\raggedbottom
	
	\title{DRIFT: Risk-Constrained Diffusion with Imitation Priors for Mixed-Autonomy Traffic Generation}
	
	\author{Yaoshen Yu, Minghui Liwang,~\IEEEmembership{Senior Member, IEEE}, Wenbo Zhu, Xinlei Yi,~\IEEEmembership{Senior Member, IEEE}, Zhang Liu, Yiguang Hong,~\IEEEmembership{Fellow, IEEE}, Xianbin Wang,~\IEEEmembership{Fellow, IEEE}, and Seyyedali Hosseinalipour,~\IEEEmembership{Senior Member, IEEE}%
		\thanks{Y. Yu (yys806@tongji.edu.cn), M. Liwang (minghuiliwang@tongji.edu.cn), W. Zhu (wbzhu@tongji.edu.cn), X. Yi (xinleiyi@tongji.edu.cn), and Y. Hong (yghong@tongji.edu.cn) are with the Department of Control Science and Engineering, the Shanghai Research Institute for Intelligent Autonomous Systems, the State Key Laboratory of Autonomous Intelligent Unmanned Systems, Tongji University, Shanghai, China. Z. Liu (zliu3224@uwo.ca) and X. Wang (xianbin.wang@uwo.ca) are with the Department of Electrical and Computer
			Engineering, Western University, London, Ontario, Canada. S. Hosseinalipour (alipour@buffalo.edu) is with the Department of Electrical Engineering, University at Buffalo-SUNY, NY, USA. Corresponding author: Minghui Liwang.}
		\vspace{-5mm}}
	
	\markboth{IEEE Transactions on Intelligent Transportation Systems}%
	{Yu \MakeLowercase{\textit{et al.}}: DRIFT for Mixed-Autonomy Traffic Generation}
	
	\maketitle
	
	\begin{abstract}
		Future intelligent transportation systems are envisioned to evolve toward a long-term mixed-autonomy paradigm, where human-driven vehicles (HVs) and autonomous vehicles (AVs) coexist within highly coupled traffic ecosystems. Such coexistence introduces pronounced heterogeneity, amplified uncertainty, and increasingly intricate interaction dynamics. In this context, it remains fundamentally challenging to simultaneously capture the heterogeneous behavioral distribution shifts arising from dynamic AV penetration, generate diverse yet executable trajectories under strong inter-vehicle coupling, and conduct reliable closed-loop safety and stability diagnostics for rare but high-impact events. To this end, we present \underline{D}iffusion with \underline{R}isk constraints, \underline{I}mitation priors, and long-tail \underline{F}eedback for mixed-autonomy \underline{T}raffic generation (DRIFT), a mixed-autonomy traffic generation framework that unifies heterogeneity-aware conditional encoding, conditional diffusion-based executable trajectory generation, and progressive adversarial alignment enhanced by risk-aware long-tail feedback, thereby enabling traffic behaviors to be iteratively generated, filtered, selected, and validated within a closed-loop execution pipeline. In addition, a unified evaluation protocol is developed to jointly characterize safety, efficiency, and closed-loop stability across representative traffic scenarios and AV penetration regimes. Experimental results demonstrate that DRIFT achieves a strong safety-efficiency trade-off in closed-loop mixed-autonomy benchmarks, while further revealing the critical influence of candidate executability, online selection, and long-tail feedback on executable traffic evolution.
	\end{abstract}
	
	\vspace{-1.5mm}

	\vspace{-1.2mm}
	\section{Introduction}\label{sec:introduction}
	
	\IEEEPARstart{T}{he} rapid advancement of artificial intelligence (AI) and automotive engineering has reshaped the research landscape of autonomous driving, where traffic-flow simulation and scenario generation have emerged as indispensable tools for developing next-generation autonomous driving systems. Accordingly, representative platforms and generators, such as Flow \cite{flow2022}, SceneDiffuser \cite{scenediffuser2024}, TrafficMCTS \cite{trafficmcts2025}, and Causal Driving Pattern Transfer (CDPT) \cite{chen2026cdpt}, enable the generation of realistic driving scenarios for technological development, benchmarking, and closed-loop evaluation. Beyond serving as experimental testbeds, these platforms provide a  framework for assessing safety, efficiency, robustness, and generalization capabilities of autonomous-driving algorithms under diverse traffic conditions, environmental factors, and operational uncertainties.
	
	Looking ahead, future intelligent transportation systems will increasingly operate in mixed-autonomy traffic environments where human-driven vehicles (HVs) and autonomous vehicles (AVs) coexist under dynamically changing AV penetration rates \cite{fang2024mixed,li2024hybrid,wang2024mixedsafe,cen2025mixedflow,wang2025energy,pei2024capacity,wang2025transfer,wang2025intersection}. The resulting traffic ecosystem is characterized by substantial behavioral heterogeneity, strong inter-vehicle coupling, and non-stationary interaction patterns, which complicate both traffic generation and evaluation. In this context, 
	% realistic microscopic behaviors are valuable only if they remain executable and induce plausible traffic evolution under closed-loop deployment. Therefore, 
	evaluating a traffic-generation framework solely through behavioral realism is insufficient. Instead, realism must be assessed jointly with executability, safety, efficiency, and traffic stability, necessitating generation models that are explicitly designed and evaluated within a closed-loop setting.
	
	Nevertheless, existing approaches address only subsets of the challenges arising in mixed-autonomy traffic. For example, rule-based and control-theoretic methods can  enforce safety and control objectives, but may struggle to scale to highly interactive multi-agent environments with complex behavioral diversity \cite{fang2024mixed,gongzhu2024,wang2025intersection}. Data-driven approaches based on behavior cloning and imitation learning improve behavioral realism, yet often regress toward average driving patterns and fail to adequately capture diverse interaction modes \cite{urbangail2024,ctx2trajgen2025}. Adversarial imitation learning-based methods further enhance the alignment between the generated and expert driving trajectories but may suffer from training instability and limited exposure to rare safety-critical events \cite{diffail2024,fastgail2025,gmail2025}. More recently, diffusion-based methods have shown strong capabilities in generating diverse traffic behaviors and trajectories; however, without explicit executability and feasibility considerations, their generated trajectories may not remain physically realizable during closed-loop deployment \cite{scenediffuser2024,controltraj2024,diffrntraj2024,zheng2025fg}. As a result, existing studies largely focus on individual aspects of realism, safety, or executability in isolation, while a framework that jointly models heterogeneous mixed-autonomy behaviors, generates executable traffic evolution, and supports risk-aware closed-loop safety evaluation remains lacking.
	
	\vspace{-3mm}
	\subsection{Research Questions and Related Work}\label{sec:core_motivation}
	Motivated by the above limitations, we formulate three  research questions (RQs) that identify the gap in the existing literature and define our central objectives.
	
	\noindent
	$\bullet$~\textit{RQ 1: How can traffic-generation models adapt to changing mixtures of HVs and AVs?}~As the proportion, or penetration, of AVs increases, the composition of traffic fundamentally changes. Specifically, traffic patterns (e.g., reaction time, car-following behavior, and lane-changing decisions) observed in predominantly human-driven environments may differ significantly from those observed when AVs constitute a large fraction of the vehicle population. 
	Consequently, a model trained under one AV penetration level may fail to generalize to another. Nevertheless, existing studies on mixed-autonomy traffic, including flow-based approaches \cite{flow2022}, game-theoretic formulations \cite{fang2024mixed}, and robust control methods \cite{wang2025intersection,gongzhu2024}, typically focus on specific AV penetration settings. Likewise, studies investigating hybrid stability, safety-aware traffic control, trust-dependent capacity, takeover-induced stability, energy efficiency, and heterogeneous platooning primarily analyze mixed-autonomy traffic under fixed operating conditions rather than explicitly modeling how traffic behavior evolves as AV penetration changes \cite{li2024hybrid,wang2024mixedsafe,pei2024capacity,wang2025transfer,cen2025mixedflow,wang2025energy,ren2024platoons}. Similarly, although data-driven methods, such as CATF \cite{contextaware2024}, interaction-aware and driving-style-aware trajectory prediction \cite{zhang2025style}, post-interactive trajectory prediction \cite{huang2025pioformer}, multi-agent imitation learning \cite{sunkim2024}, and transformer-based multi-agent reinforcement learning (MARL) \cite{li2025stfusion}, improve context and interaction modeling, they generally do not explicitly account for how traffic behavior evolves as the composition of HVs and AVs changes. Therefore, a key challenge is to develop a traffic representation that simultaneously captures vehicle-specific characteristics and the global mixed-autonomy composition of the traffic system. To address this challenge, we develop a heterogeneity-aware conditioning mechanism that jointly encodes vehicle type and AV penetration into a unified traffic representation.
	
	\noindent
	$\bullet$~\textit{RQ 2: How can generated traffic behaviors remain executable during closed-loop deployment?}~In many traffic scenarios, particularly at intersections, merges, and other conflict-prone regions, multiple driving behaviors may appear reasonable for the same traffic situation. For example, a vehicle may choose to accelerate, yield, or maintain its trajectory depending on the actions of surrounding vehicles. Due to this phenomenon, while modern generative models can produce diverse driving behavior candidates, not all generated behaviors can be safely executed: for example, a generated trajectory may appear realistic in isolation yet require infeasible accelerations, violate roadway constraints, or create unsafe interactions with neighboring vehicles. Motivated by this, recent studies, including ControlTraj \cite{controltraj2024}, Diff-RNTraj \cite{diffrntraj2024}, Diffusion-Planner \cite{zheng2025fg}, and SceneDiffuser \cite{scenediffuser2024}, have shown the potential of diffusion-based and planning-oriented generation methods for producing controllable vehicle trajectories. Also, scene-level and controllable diffusion studies, such as DiffScene \cite{diffscene2024}, SceneControl \cite{scenecontrol2024}, OGD/ECM-guided generation \cite{jointdiff2024}, and intention-aware diffusion prediction \cite{intentiondiff2024}, have improved the controllability and context-awareness of generated traffic behaviors. Similarly, intent-aware prediction and geometric refinement techniques have been  shown to improve the feasibility of generated trajectories~\cite{kang2026planningoriented}. However, these approaches generally focus on trajectory generation itself rather than ensuring that generated behaviors remain executable when interacting with other vehicles in a closed-loop  system. Thus, a key challenge is to jointly account for vehicle dynamics, roadway constraints, and inter-vehicle interactions during behavior candidate generation. To address this challenge, we develop a generative framework that produces multiple behavior candidates and evaluates their executability via feasibility-aware rollout and online candidate selection prior to execution.
	
	\noindent
	$\bullet$~\textit{RQ 3: How can rare safety-critical events be systematically incorporated into traffic generation and evaluation?}~In mixed-autonomy traffic, the most consequential failures often arise from rare but high-impact events, such as near-collisions, unsafe merges, deadlocks, and traffic instabilities that propagate through the surrounding traffic flow. 
	% Although these events occur infrequently, they can significantly affect the safety and reliability of the overall transportation system. 
	Consequently, a traffic-generation framework should not only reproduce common driving behaviors, but also evaluate and learn from safety-critical situations that emerge during closed-loop execution. To this end, adversarial imitation learning (AIL) methods, including PS-TrajGAIL \cite{urbangail2024}, DiffAIL \cite{diffail2024}, FS-GAIL \cite{fastgail2025}, and GMAIL \cite{gmail2025}, improve  the alignment between generated and expert driving
	trajectories, while TrafficMCTS \cite{trafficmcts2025}, DragTraffic \cite{dragtraffic2024}, and LD-Scene \cite{peng2026ldscene} evaluate safety-critical scenarios. Also, complementary work on dynamic test generation, edge-case synthesis, corner-case detection, and uncertainty-aware out-of-distribution safety assessment further emphasizes the need to identify rare but high-impact traffic events \cite{jia2024dynamictest,jing2025edge,schicktanz2025corner,wang2025ood}.
	However, these approaches generally treat risk discovery, behavior generation, and performance evaluation as separate processes. As a result, risks identified during execution are rarely leveraged to influence future behavior candidate-selection or traffic-level assessment. Thus, a key challenge is to establish a closed-loop mechanism that continuously identifies safety-critical behaviors, prioritizes them during candidate evaluation, and quantifies their impact on traffic performance. To address this challenge, we incorporate progressive adversarial alignment and targeted long-tail feedback into the rollout process, enabling risk discovery, candidate reweighting, and traffic-level robustness evaluation within a closed-loop framework. 
	
	\textit{$\star$ A more detailed discussion of existing works and their scope comparison with DRIFT is provided in Appendix~\ref{sec:appendix_related_work}.}
	
	\vspace{-3mm}
	\subsection{Novelty and Contribution}\label{sec:contributions}
	
	Addressing the above RQs leads to our framework termed \underline{D}iffusion with \underline{R}isk constraints, \underline{I}mitation priors, and long-tail \underline{F}eedback for mixed-autonomy \underline{T}raffic generation (DRIFT), which in turn entails the following contributions:
	% an integrated traffic-generation framework that combines heterogeneity-aware behavior modeling, executable candidate diffusion, and risk-aware closed-loop evaluation for mixed-autonomy traffic systems. The main contributions of this work can be summarized as follows:
	
	\noindent
	$\bullet$~\textit{Unified formulation for mixed-autonomy traffic generation and evaluation:} We formulate a closed-loop mixed-autonomy traffic-generation problem in which HVs and AVs coexist under varying AV penetration rates. The formulation  captures three fundamental challenges: \textit{(i)} adaptation to changing AV penetration rates (i.e., traffic composition), \textit{(ii)} generation of executable behaviors under inter-vehicle interactions, and \textit{(iii)}  evaluation of traffic performance under safety-critical events.
	
	\noindent
	$\bullet$~\textit{Diffusion-based executable traffic-generation framework:} In DRIFT, we integrate heterogeneity-aware conditional encoding (Module A), conditional diffusion-based candidate generation (Module B), and progressive adversarial alignment with long-tail feedback (Module C). Operating within a rolling-horizon framework, DRIFT generates multiple candidate behaviors, evaluates their executability, and selects actions for deployment in closed-loop traffic evolution.
	
	\noindent
	$\bullet$~\textit{Risk-aware online candidate-selection mechanism:} We develop a recurrent observation-generation-selection-execution-feedback pipeline, where candidate behaviors are evaluated according to behavioral realism, traffic efficiency, safety risk, and execution feasibility. This pipeline bridges offline generative modeling and online traffic evolution, enabling candidate-level decisions to directly influence traffic-level outcomes.
	
	\noindent
	$\bullet$~\textit{Extensive traffic-level validation:} We evaluate DRIFT across various mixed-autonomy and AV penetration settings in a closed-loop \textit{Flow}/\textit{SUMO} environment. Through baseline comparisons and ablation studies, we assess how different components of DRIFT affect safety, efficiency, and executability.

	\vspace{-2.2mm}
	\section{Key Definitions and Core Modeling}\label{sec:overview_modeling}
	\subsection{Key Definitions}\label{sec:key_definitions}
	To ground our discussion, we first enumerate the key definitions used throughout this paper (interested readers can refer to Appendix~\ref{sec:appendix_notations} for a summary of the used notations).
	
	\noindent\hangindent=\parindent\hangafter=1\makebox[\parindent][l]{$\bullet$}\textbf{Time step (TS)}: We discretize the traffic evolution into TSs indexed by $t\in\mathcal{T}=\{0,1,\dots,T\}$. Also, the interval between two adjacent TSs is denoted by $\Delta t$.
	
	\noindent\hangindent=\parindent\hangafter=1\makebox[\parindent][l]{$\bullet$}\textbf{Replanning instant (RI)}: RIs form a subset of TSs, denoted by $\mathcal{T}^{\mathsf{replan}}\subseteq\mathcal{T}$. At each RI, the current traffic state is observed and a new closed-loop planning cycle starts. Except for the initial RI, this state is the simulator state obtained after applying the previously selected controls.
	
	\noindent\hangindent=\parindent\hangafter=1\makebox[\parindent][l]{$\bullet$}\textbf{Planning window (PW)}: Given an RI $t$, we define a PW as:
	\begin{equation}
		\mathcal{T}^{\mathsf{plan}}_t=\{t+1,t+2,\dots,t+H^{\mathsf{plan}}\},
	\end{equation}
	where $\mathcal{T}^{\mathsf{plan}}_t$ represents a future window defined at RI $t$, and $H^{\mathsf{plan}}$ is the number of future TSs within each PW.
	
	\noindent\hangindent=\parindent\hangafter=1\makebox[\parindent][l]{$\bullet$}\textbf{Planning step (PS)}: A PS indicates a future TS inside PW $\mathcal{T}^{\mathsf{plan}}_t$. Specifically, the $h$-th PS corresponds to TS $t+h$, while its index satisfies $h\in\{1,\dots,H^{\mathsf{plan}}\}$.
	
	\noindent\hangindent=\parindent\hangafter=1\makebox[\parindent][l]{$\bullet$}\textbf{Execution window (EW)}: After planning, only the first $H^{\mathsf{exec}}$ PSs of the PW are executed before the next RI is reached, where $H^{\mathsf{exec}}\le H^{\mathsf{plan}}$. This defines the EW as:
	\begin{equation}
		\mathcal{T}^{\mathsf{exec}}_t=\{t+1,t+2,\dots,t+H^{\mathsf{exec}}\}.
	\end{equation}
	This design balances long-horizon interaction assessment with closed-loop adaptability by evaluating future traffic evolution over an extended horizon while executing only a short segment before incorporating new feedback.
	
\begin{figure}[t]
		\vspace{-2mm}
		\centering
		\includegraphics[width=\columnwidth]{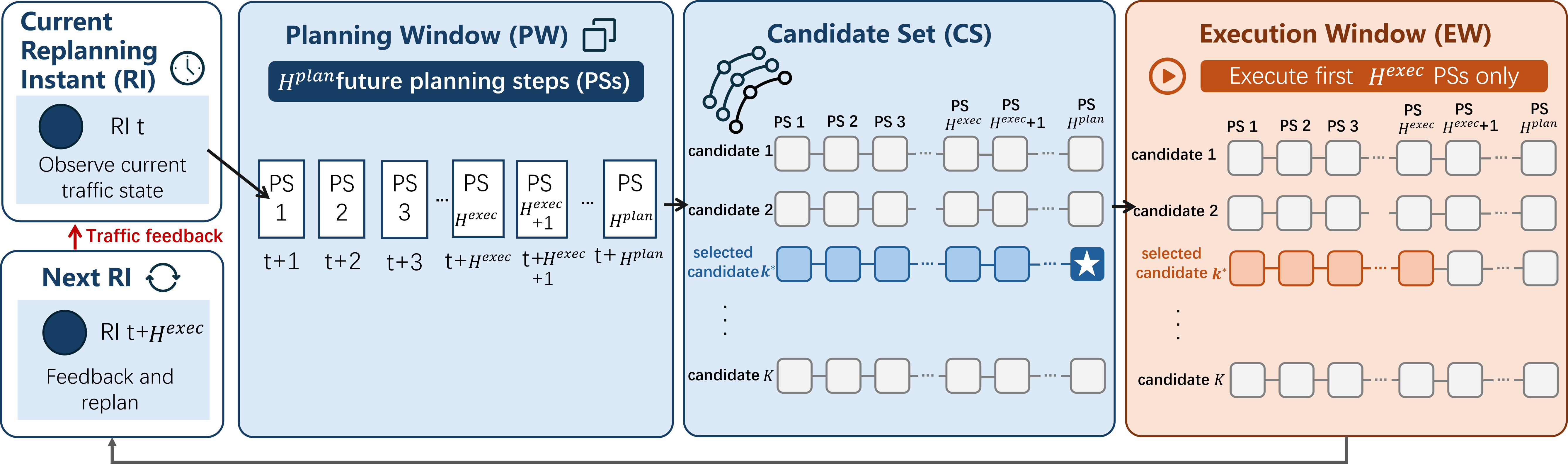}
		\vspace{-7.5mm}
		\caption{Illustration of temporal-window definitions in this work. TSs are discrete points separated by $\Delta t$. Given the current RI, we define a local PW $\mathcal{T}^{\mathsf{plan}}_t$ of future PSs, generate and rank a CS over the PW, execute only the first $H^{\mathsf{exec}}$ PSs within the EW $\mathcal{T}^{\mathsf{exec}}_t$, and then reach the next RI.}
		\label{fig:rolling_horizon_timeline}
		\vspace{-5mm}
	\end{figure}

	\noindent\hangindent=\parindent\hangafter=1\makebox[\parindent][l]{$\bullet$}\textbf{Candidate and candidate set (CS)}: A candidate consists of an executable control sequence over the PW together with its induced future trajectory. For each active vehicle at an RI, a CS that contains $K$ candidates is generated. Candidates are subsequently evaluated according to a selection rule, and the selected candidate is executed over the EW. Fig.~\ref{fig:rolling_horizon_timeline} illustrates the resulting TS--RI--PW--PS--EW relationship.
	
	\vspace{-3mm}
	\subsection{Modeling of Vehicle-Level State and Interaction}\label{sec:vehicle_state}
	Let $\mathcal{V}=\{1,\dots,i,\dots,|\mathcal{V}|\}$ denote the set of all vehicles. 
	% We next define the vehicle-level state and interaction variables used throughout the framework. 
	% To distinguish between human-driven and autonomous operation, 
	Each vehicle $i\in\mathcal{V}$ at TS $t$ is associated with a control-authority indicator $\eta_t^{i}\in\{\mathsf{HV},\mathsf{AV}\}$, where $\eta_t^{i}=\mathsf{HV}$ and $\eta_t^{i}=\mathsf{AV}$ indicate human control and autonomous driving at TS $t$, respectively. This formulation naturally accommodates mixed-autonomy traffic as well as transitions between control modes, such as takeover scenarios. For a given TS $t$, let $\tilde{\mathcal{V}}_t$ denote the set of active vehicles in the network, where the HV/AV sets and AV penetration rate $\rho_t$ are defined as follows:
	\begin{equation}
		\begin{aligned}
			\tilde{\mathcal{V}}_t^{\mathsf{AV}}&=\{i\in\tilde{\mathcal{V}}_t\mid \eta_t^{i}=\mathsf{AV}\},\\
			\tilde{\mathcal{V}}_t^{\mathsf{HV}}&=\{i\in\tilde{\mathcal{V}}_t\mid \eta_t^{i}=\mathsf{HV}\},\\
			\rho_t&={|\tilde{\mathcal{V}}_t^{\mathsf{AV}}|}/{|\tilde{\mathcal{V}}_t|}.
		\end{aligned}
	\end{equation}
	% By construction, $\tilde{\mathcal{V}}_t^{\mathsf{HV}}$ and $\tilde{\mathcal{V}}_t^{\mathsf{AV}}$ form a disjoint partition of $\tilde{\mathcal{V}}_t$ at each TS. 
	For a vehicle $i\in\tilde{\mathcal{V}}_t$, we define its \textit{state vector} as:
	\begin{equation}
		\mathbf{x}_t^{i}=[p_{x,t}^{i},p_{y,t}^{i},v_t^{i},a_t^{i},\psi_t^{i}]^{\top},
	\end{equation}
	where $p_{x,t}^{i}$ and $p_{y,t}^{i}$ are $(x,y)$ planar coordinates, $v_t^{i}$ denotes the speed, $a_t^{i}$ represents the longitudinal acceleration, and $\psi_t^{i}$ is the heading angle. To capture temporal driving patterns, we further define the \textit{ego-history sequence} as:
	\begin{equation}
		\mathbf{h}_t^{i}=\{\mathbf{x}_{t-\ell}^{i}\}_{\ell=0}^{L_h-1},
	\end{equation}
	where $L_h$ denotes the history length.
	To capture surrounding traffic, we define the \textit{neighborhood} of vehicle $i$ at TS $t$ as:
	\begin{equation}
		\mathcal{N}_t^{i}=\left\{j\in\tilde{\mathcal{V}}_t\setminus\{i\}\;\middle|\;
		\operatorname{dist}_t(i,j)\le r_n\right\},
	\end{equation}
	where $\operatorname{dist}_t(i,j)$ denotes the proximity distance, and $r_n$ is the neighborhood radius.
	Here, the screening distance $\operatorname{dist}_t(i,j)$ is used to decide whether vehicle $j$ enters the local neighborhood. To define the longitudinal safety terms, we separately define the lead vehicle candidate set, based on which the line-aligned spacing  $d_t^{i,j}$ is measured. In particular, let $\ell_t^i$ and $s_t^i$ denote the lane identifier and lane-centerline curvilinear coordinate of vehicle $i$, and let $L^j$ denote the length of vehicle $j$. The set of longitudinal lead vehicles is defined as follows:
	\begin{equation}
		\mathcal{A}_t^i=\{j\in\mathcal{N}_t^i\mid \ell_t^j=\ell_t^i,\ s_t^j>s_t^i\}.
	\end{equation}
	If $\mathcal{A}_t^i\neq\emptyset$, the interacting lead is $j_t^{\mathsf{lead}}(i)=\arg\min_{j\in\mathcal{A}_t^i}\{s_t^j-s_t^i\}$, and its lane-aligned net spacing is $d_t^{i,j_t^{\mathsf{lead}}}=s_t^{j_t^{\mathsf{lead}}}-s_t^i-L^{j_t^{\mathsf{lead}}}$. If $\mathcal{A}_t^i=\emptyset$, no longitudinal lead exists and the THW/TTC entries are assigned neutral values.
	For each neighboring vehicle $j\in\mathcal{N}_t^i$, we construct the \textit{relative interaction feature} as:
	\begin{equation}
		\hspace{-4mm}
		\resizebox{0.92\linewidth}{!}{$
			\begin{aligned}
				\Delta\mathbf{x}_t^{i,j}{=}\big[\Delta p_{x,t}^{i,j},\Delta p_{y,t}^{i,j},\Delta v_t^{i,j},\Delta a_t^{i,j},
				\Delta\psi_t^{i,j},\mathrm{THW}_t^{i,j},\mathrm{TTC}_t^{i,j}\big]^{\top}\hspace{-1mm},
			\end{aligned}
			$}\hspace{-3mm}
	\end{equation}
	where $\Delta p_{x,t}^{i,j}=p_{x,t}^{j}-p_{x,t}^{i}$, $\Delta p_{y,t}^{i,j}=p_{y,t}^{j}-p_{y,t}^{i}$, $\Delta v_t^{i,j}=v_t^{j}-v_t^{i}$, $\Delta a_t^{i,j}=a_t^{j}-a_t^{i}$, and $\Delta\psi_t^{i,j}=\psi_t^{j}-\psi_t^{i}$. Also, the longitudinal safety is represented by the time headway (THW) and time-to-collision (TTC) measures, which are defined as follows:
	\begin{equation}
		\mathrm{THW}_t^{i,j}
		=\frac{d_t^{i,j}}{\max(v_t^{i},\varepsilon)},
		\label{eq:thw}
	\end{equation}
	% and the TTC measure is defined as:
	\begin{equation}
		\mathrm{TTC}_t^{i,j}
		=\begin{cases}
			\dfrac{d_t^{i,j}}{\max(v_t^{i}-v_t^{j},\varepsilon)}, & v_t^{i}>v_t^{j},\\
			\infty, & v_t^{i}\le v_t^{j},
		\end{cases}
		\label{eq:ttc}
	\end{equation}
	where $0<\varepsilon\ll 1$ is used to avoid division by zero. In \eqref{eq:thw}--\eqref{eq:ttc}, $j$ denotes the interacting lead vehicle $j_t^{\mathsf{lead}}(i)$ rather than an arbitrary neighbor (the THW/TTC entries in $\Delta\mathbf{x}_t^{i,j}$ are computed by the above formulas when $j=j_t^{\mathsf{lead}}(i)$; for other neighboring vehicles they are set to neutral values and the interaction is represented by the relative position, motion, and orientation entries). Larger THW and TTC values in \eqref{eq:thw}--\eqref{eq:ttc} correspond to larger longitudinal safety margins. Lateral conflicts, merge interactions, and intersection-related risks are  captured later through the conflict-edge set $\mathcal{E}^{\mathsf{conf}}$ in the lane-level topology graph and the candidate-level rollout diagnostics in Sections~\ref{sec:road_topology}~and~\ref{sec:candidate_selection}.
	% , which enter the feasible candidate set and the later dynamic/topological scoring term. 
	Finally, the \textit{neighborhood interaction information} of vehicle $i$ at TS $t$ is represented by:
	\begin{equation}\label{eq:NII}
		\mathbf{n}_t^{i}=\{\Delta\mathbf{x}_t^{i,j}\}_{j\in\mathcal{N}_t^{i}}.
	\end{equation}
	
	\vspace{-1.5mm}
	\subsection{Modeling of Road Topology and Heterogeneous Local State}\label{sec:road_topology}
	The road network is modeled as a directed graph given by:
	\begin{equation}
		\mathcal{G}^{\mathsf{map}}=(\mathcal{L},\mathcal{E}^{\mathsf{conn}},\mathcal{E}^{\mathsf{conf}}),
	\end{equation}
	where $\mathcal{L}$ is the lane set, $\mathcal{E}^{\mathsf{conn}}$ indicates lane connectivity, and $\mathcal{E}^{\mathsf{conf}}$ captures potential conflict relationships among lanes: a conflict implies that two lane movements geometrically overlap, merge, cross, or otherwise require coordinated occupancy, so simultaneous vehicle motion may induce collision risk, unsafe yielding, or merge-interaction constraints. Since each vehicle interacts with a limited portion of the road network, we construct a \textit{local topological graph} for vehicle $i$ at TS $t$ as:
	\begin{equation}
		\mathcal{G}^{\mathsf{map},i}_t=\left(\mathcal{L}^{i}_t,\mathcal{E}^{\mathsf{conn},i}_t,\mathcal{E}^{\mathsf{conf},i}_t\right),
		\label{eq:local_topo_subgraph}
	\end{equation}
	where $\mathcal{L}^{i}_t\subseteq\mathcal{L}$ is the set of lanes reachable by vehicle $i$ within the planning horizon $H^{\mathsf{plan}}$, while $\mathcal{E}^{\mathsf{conn},i}_t$ and $\mathcal{E}^{\mathsf{conf},i}_t$ are the corresponding connectivity and conflict edges.
	This local graph captures the roadway context most relevant to the future motion of vehicle $i$. 
	To incorporate this information into the learning framework, we use a mapping $\mu(\cdot)$ to convert the local graph into a fixed-dimensional road-semantic representation $\mathbf{m}_t^{i}$ as:
	\begin{equation}
		\mathbf{m}_t^{i}=\mu(\mathcal{G}^{\mathsf{map},i}_t),
	\end{equation}
	which summarizes reachable lanes together with their connectivity and conflict structure.
	At the traffic level,  the \textit{collective state} of all active vehicles at TS $t$ is represented by:
	\begin{equation}
		\mathbf{X}_t=\{\mathbf{x}_t^{i}\}_{i\in\tilde{\mathcal{V}}_t},
	\end{equation}
	and the corresponding interaction relationships are described through the \textit{dynamic interaction graph} given by:
	\begin{equation}
		\mathcal{G}_t^{\mathsf{int}}=(\tilde{\mathcal{V}}_t,\mathcal{E}_t),
	\end{equation}
	where $\mathcal{E}_t=\{\langle j,i\rangle\mid j\in\mathcal{N}_t^i,\ i\in\tilde{\mathcal{V}}_t\}$ is the directed interaction-edge set. An edge $\langle j,i\rangle$ indicates that vehicle $j$ belongs to the interaction neighborhood of vehicle $i$, and its edge feature is $\Delta\mathbf{x}_t^{i,j}$. Hence, the neighbor-interaction information given by \eqref{eq:NII} (i.e.,  $\mathbf{n}_t^i=\{\Delta\mathbf{x}_t^{i,j}\}_{j\in\mathcal{N}_t^i}$) naturally captures the set of edge features incident to vehicle $i$. 
	% Unlike $\mathcal{G}^{\mathsf{map}}$ and $\mathcal{G}^{\mathsf{map},i}_t$ that characterize roadway topology, $\mathcal{G}_t^{\mathsf{int}}$ captures vehicle-to-vehicle traffic interactions.
	Using the above  quantities, we define the \textit{heterogeneous local state} of vehicle $i$ as:
	\begin{equation}\label{eq:HLS}
		\mathbf{s}_t^{i}=\Big(
		\mathbf{h}_t^{i},\mathbf{n}_t^{i},
		\mathbf{m}_t^{i},\eta_t^{i}
		\Big),
	\end{equation}
	which serves as the fundamental representation used for behavior generation and decision making in our later discussions. Specifically, the word ``heterogeneous'' refers to heterogeneity across vehicles and traffic composition: $\mathbf{s}_t^i$ contains vehicle-specific history, neighbor interactions, road semantics, and the HV/AV control-authority marker, so different vehicles and AV penetration settings induce different local state distributions.
	
	\vspace{-3mm}
	\subsection{Modeling of Candidate Representation}\label{sec:candidate_selection}
	% Having defined the heterogeneous local state and the planning/execution horizons, we next introduce the notion of a candidate behavior. 
	In DRIFT, a \textit{candidate} represents a potential future action sequence for a vehicle together with the traffic trajectory induced by executing those actions. Specifically, at an RI $t\in\mathcal{T}^{\mathsf{replan}}$, the $k$-th candidate for vehicle $i$ is represented by an executable control sequence over the PW:
	\begin{equation}
		\mathbf{U}_{t}^{i,k}=\{\mathbf{u}_{t+h-1}^{i,k}\}_{h=1}^{H^{\mathsf{plan}}},\quad k\in\{1,\dots,K\},
	\end{equation}
	where $\mathbf{u}_{t+h-1}^{i,k}$ denotes the control action applied at the $h$-th PS of candidate $k$. To evaluate a candidate, we introduce the \textit{vehicle-level transition function} $f(\mathbf{x},\mathbf{u},\mathbf{s})$, which maps the current vehicle state, control input, and local traffic context to the next predicted state. This transition model is used to roll out candidate behaviors and evaluate their feasibility prior to execution; in implementation, it serves as a candidate-rollout proxy aligned with the \textit{Flow}/\textit{SUMO} state variables.\footnote{In contrast, the \textit{scene-level transition function} $\mathcal{F}(\cdot)$, introduced later in~(\ref{eq:scene_transition}), represents the actual simulator update after a candidate has been executed.} Specifically, using this transition model, the \textit{predicted states} of a vehicle over a planning horizon are defined as follows:
	\begin{equation}
		\hspace{-4mm}
		\begin{aligned}
			\widehat{\mathbf{x}}_{t}^{i,k}&=\mathbf{x}_{t}^{i},\\
			\widehat{\mathbf{x}}_{t+h}^{i,k}
			&=f\!\left(
			\widehat{\mathbf{x}}_{t+h-1}^{i,k},
			\mathbf{u}_{t+h-1}^{i,k},
			\widehat{\mathbf{s}}_{t+h-1}^{i,k}
			\right),\quad 1\le h\le H^{\mathsf{plan}},
		\end{aligned}
		\label{eq:rollout_operator}
		\hspace{-2mm}
	\end{equation}
	where $\widehat{\mathbf{s}}_{t+h-1}^{i,k}$ is constructed similar to $\mathbf{s}_t^i$ in \eqref{eq:HLS} at each rollout step over a planning horizon. 
	Let 
	$\operatorname{Rollout}_{f}$ denote the \textit{rollout operator}, which is the finite-horizon composition of $f(\cdot)$ along the candidate control sequence.     
	This operator obtains the predicted states over the PW for each candidate $k$ as follows:
	\begin{equation}\label{eq:R20}
		\boldsymbol{\tau}_{t}^{i,k}=
		\operatorname{Rollout}_{f}\!\left(\mathbf{x}_{t}^{i},\mathbf{U}_{t}^{i,k}\right)
		=\{\widehat{\mathbf{x}}_{t+h}^{i,k}\}_{h=1}^{H^{\mathsf{plan}}}.
	\end{equation}
	Subsequently, a CS for vehicle $i$ at RI $t$ is defined as:
	\begin{equation}
		\mathcal{C}_t^i=\left\{\left(\mathbf{U}_{t}^{i,k},\boldsymbol{\tau}_{t}^{i,k}\right)\right\}_{k=1}^{K}.
		\label{eq:candidate_set}
	\end{equation}
	
	To determine whether a generated candidate can be safely executed, we introduce the topology-induced \textit{feasible trajectory set} $\Omega^i(\mathcal{G}^{\mathsf{map}})$ for vehicle $i$. A rolled-out trajectory belongs to $\Omega^i(\mathcal{G}^{\mathsf{map}})$ if every predicted state remains inside the map-admissible state set induced by the local graph, its lane transitions follow $\mathcal{E}^{\mathsf{conn},i}_t$, and its conflict-edge occupancy respects $\mathcal{E}^{\mathsf{conf},i}_t$ over the PW. Let $(\mathbf{u}^{\mathsf{min}},\mathbf{u}^{\mathsf{max}})$ and $(v^{\mathsf{min}},v^{\mathsf{max}})$ denote the admissible control and speed bounds, and let $v_{t+h}^{i,k}$ be the speed component of $\widehat{\mathbf{x}}_{t+h}^{i,k}$. 
	% For this rolled-out candidate,
	Also, let $\mathrm{THW}_{t+h}^{i,k}$ and $\mathrm{TTC}_{t+h}^{i,k}$ denote the future THW and TTC values defined as:
	\begin{equation}
		\begin{aligned}
			\mathrm{THW}_{t}^{i,k,\mathsf{min}}&=\min_{1\le h\le H^{\mathsf{plan}}} \{\mathrm{THW}_{t+h}^{i,k}\},\\
			\mathrm{TTC}_{t}^{i,k,\mathsf{min}}&=\min_{1\le h\le H^{\mathsf{plan}}}\{\mathrm{TTC}_{t+h}^{i,k}\},
		\end{aligned}
	\end{equation}
	which represent the minimum safety margins encountered by candidate $k$ during rollout. To measure executability, we define the PW-level \textit{lane-topology violation rate} as follows:
	\begin{equation}
		r_t^{i,k,\mathsf{map}}=\frac{1}{H^{\mathsf{plan}}}\sum_{h=1}^{H^{\mathsf{plan}}}
		\mathbb{I}\!\left(\widehat{\mathbf{x}}_{t+h}^{i,k}\notin\Omega_{t+h}^{\mathsf{map},i}\right),
		\label{eq:main_map_violation}
	\end{equation}
	where $\Omega_{t+h}^{\mathsf{map},i}$ is the map-admissible state set induced by $\mathcal{G}^{\mathsf{map},i}_t$ at PS $h$. Thus, $r_t^{i,k,\mathsf{map}}=0$ means that all states in $\boldsymbol{\tau}_t^{i,k}$ satisfy the map-membership, lane-connectivity, and conflict-topology requirements encoded by $\Omega^i(\mathcal{G}^{\mathsf{map}})$. Let $\mathcal{Q}_{t+h}^{i,k}$ denote the vehicles that are either the lane-aligned lead of the rolled-out candidate at PS $h$ or vehicles occupying conflict edges relative to the candidate. Also, for each $j\in\mathcal{Q}_{t+h}^{i,k}$, let $d_{t+h}^{i,j,k}$ be the signed predicted clearance, which is positive under separation, zero at contact, and negative under predicted overlap. The \textit{minimum predicted clearance} over the PW is given by:
	\begin{equation}
		d_t^{i,k,\mathsf{min}}=\min_{1\le h\le H^{\mathsf{plan}}}\ \min_{j\in\mathcal{Q}_{t+h}^{i,k}} d_{t+h}^{i,j,k},
		\label{eq:main_min_clearance}
	\end{equation}
	where $d_t^{i,k,\mathsf{min}}=\infty$ if no relevant lead or conflict-edge vehicle exists. Subsequently, we define the clearance-based \textit{collision/conflict diagnostic} as:
	\begin{equation}
		r_t^{i,k,\mathsf{col}}=
		\left[\frac{d^{\mathsf{safe}}-d_t^{i,k,\mathsf{min}}}{d^{\mathsf{safe}}}\right]_+
		+\mathbb{I}\!\left(d_t^{i,k,\mathsf{min}}\le 0\right),
		\label{eq:main_collision_penalty}
	\end{equation}
	where $d^{\mathsf{safe}}>0$ is a scenario-specific safety clearance, $[z]_+=\max(z,0)$, and $\mathbb{I}(\cdot)$ is the indicator function. The first term in \eqref{eq:main_collision_penalty} penalizes any clearance below $d^{\mathsf{safe}}$, while the indicator adds an additional penalty only for predicted contact or overlap (note that the indicator threshold is set to $0$, and not $d^{\mathsf{safe}}$). Using the admissibility and safety requirements above, we define the \textit{feasible CS} $\Omega_t^{i,\mathsf{feas}}\subseteq\mathcal{C}_t^i$, which contains candidates satisfying control limits, speed limits, roadway-topology and conflict-clearance constraints, and minimum THW/TTC safety thresholds throughout the PW (i.e., each candidate in $\Omega_t^{i,\mathsf{feas}}$ is physically executable and safety-compliant), as follows:
	\begingroup
	\begin{equation}
		\hspace{-2mm}\resizebox{0.96\linewidth}{!}{$
			\begin{aligned}
				\Omega_{t}^{i,\mathsf{feas}}
				=\Big\{\left(\mathbf{U}_{t}^{i,k},\boldsymbol{\tau}_{t}^{i,k}\right)\in\mathcal{C}_t^i\ \Big|\ &
				\mathbf{u}^{\mathsf{min}}\preceq \mathbf{u}_{t+h-1}^{i,k}\preceq \mathbf{u}^{\mathsf{max}},\ \forall h,\\
				&v^{\mathsf{min}}\le v_{t+h}^{i,k}\le v^{\mathsf{max}},\ \forall h,\\
				&\boldsymbol{\tau}_{t}^{i,k}\in\Omega^i(\mathcal{G}^{\mathsf{map}})\ \text{and}\ r_t^{i,k,\mathsf{map}}=0,\\
				&r_t^{i,k,\mathsf{col}}=0,\\
				&\mathrm{THW}_{t}^{i,k,\mathsf{min}}\ge\delta^{\mathsf{THW}},\quad
				\mathrm{TTC}_{t}^{i,k,\mathsf{min}}\ge\delta^{\mathsf{TTC}}
				\Big\}.
			\end{aligned}
			$}\hspace{-5mm}
		\label{eq:omega_feas}
	\end{equation}
	\endgroup
	where $\delta^{\mathsf{THW}}$ and $\delta^{\mathsf{TTC}}$ are the longitudinal safety thresholds, and $\preceq$ denotes element-wise vector inequality.
	
	% By construction, $\Omega_{t}^{i,\mathsf{feas}}\subseteq\mathcal{C}_t^i$. 
	
	% The resulting feasible CS is then used under closed-loop dynamics specified later in Section~\ref{sec:closed_loop_replanning}.
	
	\vspace{-3mm}
	\subsection{Modeling of Closed-Loop Dynamics and State Transitions}\label{sec:closed_loop_replanning}
	Given the candidate representation above, in the following, we model how the selected candidate induces closed-loop traffic evolution. 
	% At each RI $t$, every active vehicle $i$ is assigned an execution index $k_t^{i,*}$ by the online selector applied to the feasible candidate set introduced in Section~\ref{sec:candidate_selection}. Equivalently, after the candidate scores $J_t^{i,k}$ are computed later in~\eqref{eq:score_j}, this index is obtained as $k_t^{i,*}=\arg\max_{k\in\mathcal{K}_{t}^{i,\mathsf{feas}}}J_t^{i,k}$ whenever the feasible index set is nonempty.}
	% At each RI, a CS is generated and evaluated for every active vehicle using the rolled-out trajectories, feasibility conditions, and safety/topology diagnostics defined in Section~\ref{sec:candidate_selection}. After candidate selection, only the portion of the selected candidate contained within the EW is applied to the environment, while the remaining planning horizon is used solely for future-state assessment during candidate evaluation.
	At the vehicle level, the state sequence over the EW is represented with the transition function $f(\cdot)$ used in \eqref{eq:rollout_operator} as:
	\begin{equation}
		\mathbf{x}_{t+h}^{i}=f\!\left(
		\mathbf{x}_{t+h-1}^{i},\mathbf{u}_{t+h-1}^{i},\mathbf{s}_{t+h-1}^{i}
		\right),~~ 1\le h\le H^{\mathsf{exec}},
		\label{eq:state_transition_one_step}
	\end{equation}
	% This relation records the per-vehicle state sequence of the executed segment. 
	We subsequently define the 
	% \textit{scene-level transition function} $\mathcal{F}(\cdot)$ as the 
	simulator-based \textit{closed-loop environment state update} as follows:
	\begin{equation}
		\mathbf{X}_{t+H^{\mathsf{exec}}}{=}
		\mathcal{F}(\mathbf{X}_t,\mathbf{U}_t^{\mathsf{exec}},\mathcal{G}^{\mathsf{map}}).
		\label{eq:scene_transition}
	\end{equation}
	Here, $\mathbf{U}_t^{\mathsf{exec}}{=}\{\mathbf{u}_{t+h-1}^{i}\mid i\in\tilde{\mathcal{V}}_t,\,1\le h\le H^{\mathsf{exec}}\}$ denotes the executed controls  during the EW (applying these controls for $h=1,\ldots,H^{\mathsf{exec}}$ gives the executed short segment of each  candidate). Also, the mapping $\mathcal{F}(\cdot)$ is the \textit{scene-level simulator transition}: starting from the full traffic state $\mathbf{X}_t$, it applies the vehicle-level updates in~\eqref{eq:state_transition_one_step} for the selected controls in $\mathbf{U}_t^{\mathsf{exec}}$, enforces the road-network constraints encoded by $\mathcal{G}^{\mathsf{map}}$, and returns the synchronized traffic state after the $H^{\mathsf{exec}}$ executed PSs, which is why the output state is indexed by $t+H^{\mathsf{exec}}$ in~\eqref{eq:scene_transition}. By repeatedly performing traffic-state observation, candidate generation, candidate-selection, execution over the EW, and environment feedback at successive RIs, the framework produces a complete \textit{closed-loop traffic rollout} denoted by:
	\begin{equation}
		\zeta=
		\left\{
		\left(\mathbf{X}_t,\{\mathcal{C}_t^i\}_{i\in\tilde{\mathcal{V}}_t},
		\mathbf{k}_t^{*}, \mathbf{U}_t^{\mathsf{exec}},
		\mathbf{X}_{t+H^{\mathsf{exec}}}\right)
		\right\}_{t\in\mathcal{T}^{\mathsf{replan}}},
		\label{eq:rollout_zeta}
	\end{equation}
	where $\mathbf{k}_t^{*}=\{k_t^{i,*}\}_{i\in\tilde{\mathcal{V}}_t}$ collects the selected candidate indices for active vehicles at RI $t$, with each $k_t^{i,*}$ obtained by the local candidate-selection rule via our problem formulation below.
	
	\section{Problem Formulation and Decomposition}\label{sec:problem_formulation}
	Building on the modeling in Section~\ref{sec:overview_modeling}, we next formulate our core closed-loop traffic generation problem and then decompose it into tractable subproblems to obtain its solution.
	\subsection{Core Closed-Loop Generation Problem}\label{sec:optimization_formulation}
	% The objective of DRIFT is to generate mixed-autonomy traffic rollouts that simultaneously exhibit realistic driving behaviors, maintain traffic efficiency, satisfy safety requirements, and remain physically executable during closed-loop deployment. 

	% As described in Section~\ref{sec:closed_loop_replanning}, traffic evolution is modeled through a receding-horizon process in which candidate behaviors are generated, filtered for feasibility, selected for execution, and subsequently incorporated into the next replanning cycle through environment feedback. Consequently, the quality of the resulting traffic rollout depends on (i) how traffic context is represented, (ii) how candidate behaviors are produced, and (iii) how well the generated behaviors are aligned with realistic and safety-critical traffic patterns. To capture these three functionalities, DRIFT consists of three trainable modules. Module A learns a heterogeneous traffic representation by encoding vehicle history, neighboring interactions, roadway semantics, and control authority. Module B generates candidate control sequences conditioned on the representation learned by Module A. Module C performs risk-aware distribution alignment by encouraging generated behaviors to remain consistent with expert traffic patterns while emphasizing safety-critical and long-tail traffic events. 
	
	Let $\Theta$ denote the trainable model-parameter collection of DRIFT (in essence, $\Theta=\{\phi_A,\phi_B,\vartheta\}$, where $\phi_A$ and $\phi_B$ parameterize Modules A and B, and $\vartheta$ is the trainable discriminator parameter block inside Module C; the Module-C shorthand $\phi_C$ used later also includes the non-trainable warm-up and cached reweighting quantities described in Section~\ref{sec:module_c}).     
	% For each vehicle $i$ at RI $t$, let $k_t^{i,*}$ denote the candidate selected for execution. 
	let $\mathbf{k}=\{k_t^{i,*}\mid t\in\mathcal{T}^{\mathsf{replan}},\,i\in\tilde{\mathcal{V}}_t\}$ collect the rollout-level sequence of selected  candidate indices, which determines the resulting closed-loop traffic rollout. 
	% Given a parameter set $\Theta$ and a candidate-selection policy whose realized rollout-level sequence is $\mathbf{k}$, 
	We denote the induced distribution of closed-loop traffic rollouts by $\mathbb{P}_{\Theta,\mathbf{k}}^{\mathsf{roll}}$ and the expert driving-behavior distribution, which is used as a realism reference, by $\mathbb{P}^{\mathsf{exp}}$.
	Also, to evaluate candidate quality during rollout, we assign each candidate $k$ of vehicle $i$ a composite score:
	\begin{equation}
		\begin{aligned}
			J_t^{i,k}={}\omega^{\mathsf{sim}}S_t^{i,k}
			+\omega^{\mathsf{eff}}E_t^{i,k}
			&-\omega^{\mathsf{risk}}R_t^{i,k}
			-\omega^{\mathsf{dyn}}D_t^{i,k},
		\end{aligned}
		\label{eq:score_j}
	\end{equation}
	where $S_t^{i,k}$, $E_t^{i,k}$, $R_t^{i,k}$, and $D_t^{i,k}$ quantify behavioral consistency with expert driving patterns, traffic-efficiency utility, safety risk, and execution difficulty, respectively, and their construction is detailed in Appendix~\ref{sec:appendix_scores}. In~\eqref{eq:score_j}, the coefficients  $\omega^{\mathsf{sim}},\omega^{\mathsf{eff}},\omega^{\mathsf{risk}},\omega^{\mathsf{dyn}}\geq 0$ control the relative importance of the score components. Based on these definitions, we formulate our core optimization $\mathcal{P}_0$ \textit{that seeks to maximize the cumulative quality of the selected candidates while simultaneously encouraging agreement with expert driving behaviors and emphasizing safety-critical traffic situations} as follows:
	\vspace{-3mm}
	
	\begingroup
	\small
	\setlength{\abovedisplayskip}{4pt plus 1pt minus 1pt}
	\setlength{\belowdisplayskip}{4pt plus 1pt minus 1pt}
	\setlength{\abovedisplayshortskip}{3pt plus 1pt minus 1pt}
	\setlength{\belowdisplayshortskip}{3pt plus 1pt minus 1pt}
	\setlength{\jot}{1pt}
	\begin{align}
		&\hspace{-8mm}\mathcal{P}_0:
		\max_{\Theta,\,\mathbf{k}}~
		\resizebox{0.77\columnwidth}{!}{$\displaystyle
			\mathbb{E}_{\zeta\sim\mathbb{P}_{\Theta,\mathbf{k}}^{\mathsf{roll}}}
			\!\left[
			\sum_{t\in\mathcal{T}^{\mathsf{replan}}}
			\sum_{i\in\tilde{\mathcal{V}}_t}
			\gamma^t J_t^{i,k_t^{i,*}}
			\right]
			{-}\beta^{\mathsf{real}}\Delta^{\mathsf{real}}
			{-}\beta^{\mathsf{tail}}\Delta^{\mathsf{tail}}
			$}, \label{eq:p0} \hspace{-3mm}\\[-1pt]
		\text{s.t.}~
		&k_t^{i,*}\in\{1,\dots,K\}, \tag{\ref*{eq:p0}a}\label{eq:p0index}\\[-1pt]
		&\big(\mathbf{U}_{t}^{i,k_t^{i,*}},\boldsymbol{\tau}_{t}^{i,k_t^{i,*}}\big)
		\in\Omega_t^{i,\mathsf{feas}}, \tag{\ref*{eq:p0}b}\label{eq:p0sel}\\[-1pt]
		&\mathbf{x}_{t+h}^{i}=f\!\big(\mathbf{x}_{t+h-1}^{i},
		\mathbf{u}_{t+h-1}^{i,k_t^{i,*}},\mathbf{s}_{t+h-1}^{i}\big), ~1\le h\le H^{\mathsf{exec}}, \tag{\ref*{eq:p0}c}\label{eq:p0dyn}\\
		&\mathbf{X}_{t+H^{\mathsf{exec}}}=
		\mathcal{F}\!\left(\mathbf{X}_t,\mathbf{U}_{t}^{\mathsf{exec},*},\mathcal{G}^{\mathsf{map}}\right). \tag{\ref*{eq:p0}d}\label{eq:p0scene}
	\end{align}
	\endgroup
	
	\noindent In $\mathcal{P}_0$, the variables are $\Theta$ and $\mathbf{k}$, which collect the trainable parameters and the selected candidate indices. In \eqref{eq:p0}, $\gamma\in(0,1]$ denotes the discount factor of contributions across successive RIs. The constraints are imposed for every $t\in\mathcal{T}^{\mathsf{replan}}$ and $i\in\tilde{\mathcal{V}}_t$, and for every $h\in\{1,\ldots,H^{\mathsf{exec}}\}$ where applicable.  
	The two target penalties in \eqref{eq:p0} are defined as:
	\begin{align}
		\Delta^{\mathsf{real}}
		&=\left[
		d^{\mathsf{real}}\!\left(\mathbb{P}_{\Theta,\mathbf{k}}^{\mathsf{roll}},
		\mathbb{P}^{\mathsf{exp}}\right)-\epsilon^{\mathsf{real}}
		\right]_+, \label{eq:p0real}\\
		\Delta^{\mathsf{tail}}
		&=\left[
		\operatorname{CVaR}_{\alpha}\!\left[R^{\mathsf{tail}}(\zeta)\right]
		-\epsilon^{\mathsf{tail}}
		\right]_+, \label{eq:p0tail}
	\end{align}
	where $[z]_+=\max(z,0)$, and coefficients $\beta^{\mathsf{real}},\beta^{\mathsf{tail}}\ge 0$ balance rollout distribution alignment and tail-risk reduction. In~\eqref{eq:p0real}, $d^{\mathsf{real}}(\cdot,\cdot)$ denotes the realism discrepancy induced by the finite-sample adversarial alignment loss (later defined in~\eqref{eq:BaseAlignDef}), i.e., by the ability of the state-conditioned discriminator to separate generated rollout segments from expert rollout segments under matched local states. Also, the scalar $\epsilon^{\mathsf{real}}$ captures the tolerable level of this discrepancy. In \eqref{eq:p0tail}, $R^{\mathsf{tail}}(\zeta)$ aggregates rollout-level rare-event risk over the complete closed-loop rollout $\zeta$, including safety-risk accumulation, execution violations, and severe events such as collision, deadlock, or unstable traffic degradation; $\operatorname{CVaR}_{\alpha}(\cdot)$ denotes conditional value-at-risk at confidence level $\alpha$ \cite{rockafellar2000cvar}; and $\epsilon^{\mathsf{tail}}$ denotes the tolerable level of tail-risk. CVaR is used here as a system-level tail-risk target: it emphasizes the upper-risk portion of the rollout-risk distribution rather than the average risk alone. In our implemented finite-sample solution, this target is approximated through candidate-level risk/violation diagnostics and the long-tail reweighting coefficients as described in Section~\ref{sec:module_c}. Collectively,  the target penalties \eqref{eq:p0real} and \eqref{eq:p0tail}, which are integrated within the objective in \eqref{eq:p0}, serve as soft rollout-level objectives for distribution alignment and tail-risk control. 
	
	Further, in $\mathcal{P}_0$, constraint \eqref{eq:p0index} specifies that each selected index must be one of the generated candidates, and 
	\eqref{eq:p0sel} requires the selected candidate to belong to the feasible set defined in Section~\ref{sec:candidate_selection}. Constraints \eqref{eq:p0dyn} and \eqref{eq:p0scene} impose vehicle-level and scene-level closed-loop state transitions. Particularly, in \eqref{eq:p0scene}, $\mathbf{U}_{t}^{\mathsf{exec},*}=\{\mathbf{u}_{t+h-1}^{i,k_t^{i,*}}\mid i\in\tilde{\mathcal{V}}_t,\,1\le h\le H^{\mathsf{exec}}\}$ collects the selected controls executed within the EW.

	% They guide the finite-sample alignment loss and long-tail feedback introduced later, but they should not be read as closed-form safety guarantees. By contrast, constraints \eqref{eq:p0index}--\eqref{eq:p0scene} define the operational requirements for selection, feasibility, vehicle-level rollout, and scene-level simulator transition. These constraints apply at every RI $t\in\mathcal{T}^{\mathsf{replan}}$ and for every active vehicle $i\in\tilde{\mathcal{V}}_t$, so $k_t^{i,*}$ denotes the selected execution index for that specific RI--vehicle pair.
	
	\begin{figure}[t]
		\vspace{-3mm}
		\centering
		\includegraphics[width=\columnwidth]{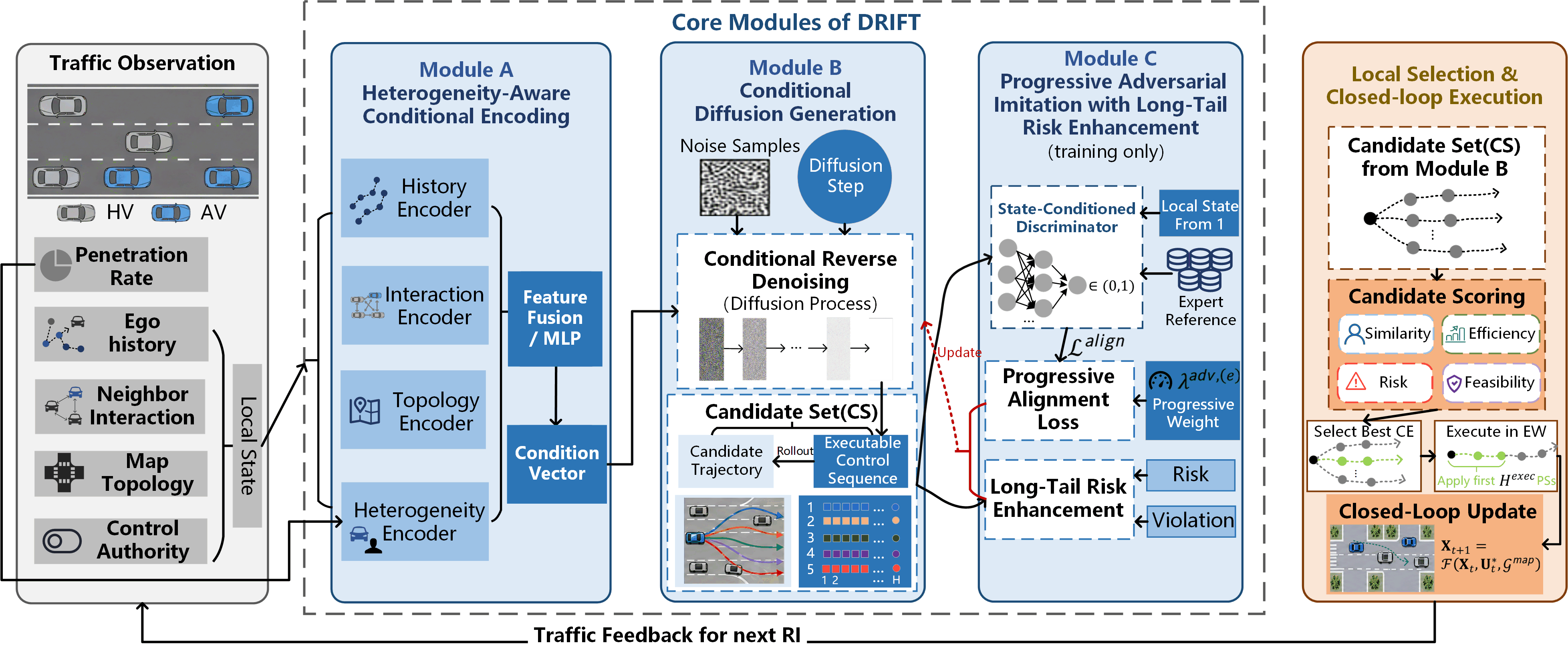}
		\vspace{-7mm}
		\caption{Overall workflow of DRIFT, separating offline training from online closed-loop rollout. During offline training, Module A encodes heterogeneous local traffic states into condition vectors, Module B learns executable candidate generation, Module C provides distribution alignment and long-tail feedback. During online rollout, trained modules are fixed; feasible candidates are scored, selected, executed within the EW, and fed back to the next RI.}
		\label{fig:overall_framework}
		\vspace{-4mm}
	\end{figure}

	\vspace{-2.5mm}
	\subsection{Problem Decomposition}\label{sec:problem_decomposition}
	
	% \noindent
	% \textit{Hardness of $\mathcal{P}_0$}.
	Problem $\mathcal{P}_0$ serves as a system-level formulation that captures the overall objectives of DRIFT. 
	% However, these objectives are evaluated at different granularities: candidate-selection decisions are made locally for individual vehicles at each RI, whereas realism is assessed over complete traffic rollouts and tail-risk is measured through the occurrence of rare safety-critical events across rollouts. Consequently, the optimization variables and objectives are inherently coupled across both local decision-making and long-horizon traffic evolution. This coupling makes the derivation of the solution of $\mathcal{P}_0$ highly complex: it would 
	% Specifically, the candidate-selection decisions depend on generated candidate sets and real-time traffic feedback, while the realism and tail-risk terms are defined through rollout-level and distribution-level statistics that are estimated from finite training samples. As a result, directly solving $\mathcal{P}_0$ 
	Nevertheless, directly solving $\mathcal{P}_0$ is highly non-trivial: it would require jointly optimizing continuous model parameters, discrete candidate-selection decisions, simulator-driven state transitions, and rare-event safety objectives within a single stochastic optimization framework.
	% \noindent
	% \textit{Problem decomposition}.~
	To address this complexity, we decompose $\mathcal{P}_0$ into a sequence of subproblems that form the operational workflow of DRIFT. Specifically, DRIFT consists of an offline training stage and an online closed-loop inference stage. \textit{During offline training}, three complementary subproblems, i.e., $\mathcal{P}_1$, $\mathcal{P}_2$, and $\mathcal{P}_3$ (separately solved via Module A, B, and C of DRIFT), are used to learn the parameter blocks in $\Theta$.   At a high level, $\mathcal{P}_1$ learns a \textit{heterogeneity-aware representation} that captures vehicle history, traffic interactions, roadway context, and control authority; $\mathcal{P}_2$ learns the \textit{conditional candidate-generation mechanism} responsible for producing future control sequences and their associated rollouts; and $\mathcal{P}_3$ performs \textit{risk-aware distribution alignment} by refining the generated behaviors using expert-driving information and long-tail safety feedback. \textit{During online closed-loop execution}, at each RI, a finite CS is generated, infeasible candidates are removed through feasibility filtering, and the online selection problem $\mathcal{P}_{\mathsf{sel}}$ selects the executed candidate, i.e., $k_t^{i,*}$. The selected candidate is then applied to the environment, producing the next traffic state and initiating the subsequent replanning cycle. 
	% Together, these subproblems shape the induced rollout distribution $\mathbb{P}_{\Theta,\mathbf{k}}^{\mathsf{roll}}$ and approximate the representation, generation, selection, and robustness requirements encoded in $\mathcal{P}_0$. 
	The DRIFT module workflow and online execution are summarized in Fig.~\ref{fig:overall_framework}.
	In the following, we introduce the above subproblems and present their solutions in Section~\ref{sec:design_drift}. 
	
	We begin by describing the offline representation-learning stage. The objective of $\mathcal{P}_1$ is to learn a traffic representation $\mathbf{c}_t^i=g_{\phi_A}(\mathbf{s}_t^i)$ through a learning pipeline parameterized by $\phi_A$, which encodes vehicle history, neighboring interactions, roadway semantics, control authority, and penetration-related information into a compact conditional representation. To this end, we formulate:
	\begin{equation}
		\mathcal{P}_1:\quad
		\min_{\phi_A}\;\mathcal{L}^{\mathsf{repr}}(\phi_A).
		\label{eq:p1}
	\end{equation}
	% Unlike the subsequent subproblems, $\mathcal{P}_1$ does not generate candidate CSs or traffic rollouts. Instead, 
	We note that $\mathcal{P}_1$ focuses solely on learning informative representations that capture the heterogeneous characteristics of mixed-autonomy traffic. Consequently, execution-related constraints of $\mathcal{P}_0$ are not imposed at this stage. Rather, the quality of the learned representation is governed through the representation-learning objective $\mathcal{L}^{\mathsf{repr}}$ detailed in Section~\ref{sec:design_drift}.
	
	% We next consider the candidate-generation stage. 
	Building upon the representation learned by $\mathcal{P}_1$, the objective of $\mathcal{P}_2$ is to generate diverse candidate control sequences and their corresponding future traffic trajectories through a learning pipeline parameterized by $\phi_B$. To this end, we formulate:
	\begin{align}
		\mathcal{P}_2:\quad
		\min_{\phi_B}\quad
		&\mathcal{L}^{\mathsf{gen}}(\phi_B)
		\label{eq:p2}\\
		\text{s.t.}\quad
		&\eqref{eq:p0sel},\ \eqref{eq:p0dyn}. \tag{\ref*{eq:p2}a}\label{eq:p2a}
	\end{align}
	The constraint in \eqref{eq:p2a} captures the feasibility and dynamic-transition requirements of the generated candidates. We note that, unlike the later formulated $\mathcal{P}_{\mathsf{sel}}$ that operates during online rollout, $\mathcal{P}_2$ is trained offline and does not determine which candidate will ultimately be executed. Instead, its role is to learn a generation mechanism that produces candidate behaviors that are diverse, dynamically consistent, and likely to satisfy the feasibility constraints. Accordingly, the requirements in \eqref{eq:p0sel} and \eqref{eq:p0dyn} are enforced through rollout-based training objectives defined in $\mathcal{L}^{\mathsf{gen}}$, which is later concretized in Section~\ref{sec:design_drift}.
	Also, the remaining operational constraints in $\mathcal{P}_0$ are handled online: the index constraint~\eqref{eq:p0index} is enforced by $\mathcal{P}_{\mathsf{sel}}$, later defined in \eqref{eq:psel}, as it selects from the finite index set $\{1,\ldots,K\}$, and the scene-transition constraint~\eqref{eq:p0scene} is enforced during EW execution by the simulator transition $\mathcal{F}(\cdot)$ defined in~\eqref{eq:scene_transition}.
	
	Finally, we address the distribution alignment stage, noting that while $\mathcal{P}_2$ learns to generate feasible and diverse candidate behaviors, it does not explicitly ensure that the resulting traffic rollouts remain aligned with expert driving patterns or adequately emphasize rare safety-critical events. To address these objectives, we formulate:
	\begin{equation}
		\mathcal{P}_3:\quad
		\min_{\vartheta}\;
		\mathcal{L}^{\mathsf{align},(e)}(\vartheta;\phi_B^{\mathsf{fix}}).
		\label{eq:p3}
	\end{equation}
	% Equation~\eqref{eq:p3} is the finite-sample training objective used for the rollout-level alignment term in \eqref{eq:p0real}. 
	% For each training pair $(t,i)$, Module C compares the expert trajectory $\boldsymbol{\tau}_{t}^{i,\mathrm{exp}}$ with the generated candidates $\{\widehat{\boldsymbol{\tau}}_{t}^{i,k}\}_{k=1}^{K}$ through the alignment loss $\mathcal{L}^{\mathsf{align},(e)}$. The long-tail part is handled separately through candidate-level diagnostics. Specifically, each generated candidate has a safety-risk score $R_t^{i,k}$ and a dynamic/topological violation score $D_t^{i,k}$. Candidates with larger $R_t^{i,k}+D_t^{i,k}$ are assigned larger reweighting coefficients in Module C and therefore receive larger weights in the next Module-B generator update. In this way, 
	Here, $\phi_B^{\mathsf{fix}}$ denotes the fixed/final Module-B generator used to produce the candidate trajectories evaluated by Module C; it appears after the semicolon in $\mathcal{L}^{\mathsf{align},(e)}(\vartheta;\phi_B^{\mathsf{fix}})$ only to indicate this fixed data-generation source, while the optimized variable in $\mathcal{P}_3$ is the discriminator parameter block $\vartheta$. As concretized in Section~\ref{sec:design_drift}, the distribution alignment term in \eqref{eq:p0real} is implemented by $\mathcal{L}^{\mathsf{align},(e)}$, while the tail-risk term in \eqref{eq:p0tail} is approximated by deterministic reweighting of finite generated candidates according to their candidate-level risk and violation scores. Thus, $\mathcal{P}_3$ approximates the CVaR-oriented tail defined in \eqref{eq:p0tail} through an empirical upper-tail surrogate: candidates with large $R_t^{i,k}$ and $D_t^{i,k}$ receive larger weights in subsequent generator updates, so finite-sample training identifies candidate segments that are most likely to contribute to high rollout risk.
	
	During online closed-loop execution, the execution index is determined locally for each RI-vehicle pair. In doing so, we first note that for each active vehicle $i$ at RI $t$, Module B generates the finite CS $\mathcal{C}_t^i=\{(\mathbf{U}_{t}^{i,k},\boldsymbol{\tau}_{t}^{i,k})\}_{k=1}^{K}$ defined in~\eqref{eq:candidate_set}, and infeasible candidates are removed according to $\Omega_t^{i,\mathsf{feas}}$. This yields the feasible candidate-index set:
	\begin{equation}
		\mathcal{K}_{t}^{i,\mathsf{feas}}
		=\{k\mid(\mathbf{U}_{t}^{i,k},\boldsymbol{\tau}_{t}^{i,k})\in\Omega_t^{i,\mathsf{feas}}\}.
	\end{equation}
	Since $\mathcal{K}_{t}^{i,\mathsf{feas}}$ is finite, the online local selection problem can be solved by direct enumeration. Specifically, for a given RI-vehicle pair $(t,i)$,
	% , let us write $k^*$ as a local shorthand for $k_t^{i,*}$: 
	each feasible candidate is scored according to the value of (\ref{eq:score_j}), and the candidate with the largest score is executed. This can be formulated as follows:
	\begin{align}
		\mathcal{P}_{\mathsf{sel}}:\quad
		k_t^{i,*}
		&=\arg\max_{k}J_t^{i,k}
		\label{eq:psel}\\
		\text{s.t.}\quad
		&k\in\mathcal{K}_{t}^{i,\mathsf{feas}}. \tag{\ref*{eq:psel}a}\label{eq:psel_feas}
	\end{align}
	We note that no learnable parameter is updated in $\mathcal{P}_{\mathsf{sel}}$; its input is the generated CS, the feasible subset when nonempty, and the score values, and its output is only the executed index $k^*$ for the current RI-vehicle pair. If $\mathcal{K}_{t}^{i,\mathsf{feas}}=\emptyset$, all generated candidates violate at least one condition in $\Omega_t^{i,\mathsf{feas}}$. In this rare case, the online controller applies a deterministic \textit{fallback} rather than terminating the rollout. The fallback first forms a hard-admissible subset containing candidates that still satisfy the simulator-executable conditions: control bounds, speed bounds, map/topology consistency ($r_t^{i,k,\mathsf{map}}=0$), and predicted noncollision/conflict clearance ($r_t^{i,k,\mathsf{col}}=0$). 
	% These conditions are called hard because violating them can make the selected segment nonexecutable in the simulator. 
	If this hard-admissible subset is nonempty, the fallback ignores the realism and efficiency rewards $S_t^{i,k}$ and $E_t^{i,k}$ and selects the candidate with the smallest risk--violation score $\omega^{\mathsf{risk}}R_t^{i,k}+\omega^{\mathsf{dyn}}D_t^{i,k}$. If the hard-admissible subset is empty, DRIFT delegates the current EW to the fail-safe action returned by the \textit{Flow}/\textit{SUMO} interface; this action is computed by the simulator from the current traffic state using its car-following and collision-avoidance rules.
	% This fallback is used only to preserve closed-loop executability under rare infeasible candidate sets and is not treated as an additional learnable module. 
	% Together, $\mathcal{P}_1$--$\mathcal{P}_3$ make the core closed-loop problem trainable, while $\mathcal{P}_{\mathsf{sel}}$ handles online candidate execution.

	\section{Design of DRIFT}\label{sec:design_drift}
	
	%\subsection{Framework Overview}\label{sec:framework_overview}
	
	We next detail the components of DRIFT that solve the subproblems in Section~\ref{sec:problem_decomposition} (the pseudo-code of DRIFT, which summarizes its steps, is presented in Appendix~\ref{sec:appendix_workflow}). 
	
	\subsection{Module A: Heterogeneity-Aware Condition Encoding}\label{sec:module_a}
	
	Module A solves $\mathcal{P}_1$ by mapping the heterogeneous local state $\mathbf{s}_t^{i}$ into a compact condition vector $\mathbf{c}_t^{i}$ for subsequent candidate generation. Specifically, Module A entails a set of encoders that process the four sources defined in Section~\ref{sec:road_topology} (i.e., ego-history sequence, neighborhood interaction information, road-semantic representation, and mixed-autonomy attributes):
	\begin{equation}
		\begin{aligned}
			\mathbf{z}_{h,t}^{i}&=\varphi_h(\mathbf{h}_t^{i}),\quad \quad
			\mathbf{z}_{n,t}^{i}=\sum_{j\in\mathcal{N}_t^{i}}\alpha_t^{i,j}\,\varphi_n(\Delta\mathbf{x}_t^{i,j}),\\
			\mathbf{z}_{m,t}^{i}&=\varphi_m(\mathbf{m}_t^{i}), \quad \quad
			\mathbf{z}_{r,t}^{i}=\mathbf{e}_{\eta}(\eta_t^{i})+\mathbf{e}_{\rho}(\rho_t),
		\end{aligned}
	\end{equation}
	where the encoders $\varphi_h(\cdot)$, $\varphi_n(\cdot)$, and $\varphi_m(\cdot)$ map their respective inputs into vector representations, and $\alpha_t^{i,j}$ is a normalized interaction-attention weight that reflects the relative importance of neighboring vehicle $j$. The above representations are then concatenated and transformed into a unified condition vector:
	\begin{equation}
		g_{\phi_A}(\mathbf{s}_t^{i})=
		\mathrm{MLP}\!\left([\mathbf{z}_{h,t}^{i};\mathbf{z}_{n,t}^{i};\mathbf{z}_{m,t}^{i};\mathbf{z}_{r,t}^{i}]\right),
	\end{equation}
	where MLP denotes a multilayer perceptron. The output $\mathbf{c}_t^{i}=g_{\phi_A}(\mathbf{s}_t^{i})$ serves as the condition representation passed to Module B. To make the learned representation sensitive to mixed-autonomy heterogeneity rather than only local vehicle motion, each RI-vehicle training sample $(t,i)$ is assigned a coarse behavioral-interaction tag $b_t^i$, constructed from the vehicle's control authority, short-horizon motion characteristics, and local interaction patterns. This tag is used only during representation learning to identify similar and dissimilar training samples as:
	\begin{equation}
		\begin{aligned}
			(t',j)\in\mathcal{S}_t^{i,+}&\iff b_{t'}^j=b_t^i,\\
			(t',j)\in\mathcal{S}_t^{i,-}&\iff b_{t'}^j\neq b_t^i,
		\end{aligned}
	\end{equation}
	where $(t',j)$ is drawn from the same training pool. Module A is trained using the objective:
	\begin{equation}
		\mathcal{L}^{\mathsf{repr}}(\phi_A)=\mathcal{L}^{\mathsf{stb}}(\phi_A)
		+\lambda^{\mathsf{sep}}\mathcal{L}^{\mathsf{sep}}(\phi_A)
		+\lambda^{\mathsf{reg}}\|\phi_A\|_2^2,
	\end{equation}
	\begin{equation}
		\begin{aligned}
			\mathcal{L}^{\mathsf{stb}}(\phi_A)&=\frac{1}{|\mathcal{S}_t^{i,+}|}
			\sum_{(t',j)\in\mathcal{S}_t^{i,+}}\|\mathbf{c}_t^{i}-\mathbf{c}_{t'}^{j}\|_2^2,\\
			\mathcal{L}^{\mathsf{sep}}(\phi_A)&=\frac{1}{|\mathcal{S}_t^{i,-}|}
			\sum_{(t',j)\in\mathcal{S}_t^{i,-}} \hspace{-3mm}
			\Big(\big[m^{\mathsf{sep}}-\|\mathbf{c}_t^{i}-\mathbf{c}_{t'}^{j}\|_2\big]_+\Big)^2\hspace{-1.5mm},
		\end{aligned}
		\vspace{-1mm}
	\end{equation}
	where $\mathcal{L}^{\mathsf{stb}}$ encourages samples with similar behavioral-interaction tags to have similar condition representations, while $\mathcal{L}^{\mathsf{sep}}$ encourages representations associated with different tags to remain distinguishable under margin $m^{\mathsf{sep}}$. Overall, Module A provides an AV penetration- and vehicle interaction-aware condition prior that serves as the input to Module B in Section~\ref{sec:module_b}. Its training procedure is summarized in Algorithm~\ref{alg:module_a} (the stopping criteria used in the last line of Algorithms~\ref{alg:module_a}, and the later introduced Algorithms~\ref{alg:module_b} and \ref{alg:module_c}, implies that training stops when the prescribed maximum epoch budget is reached or the validation loss does not improve).
	
	\begin{algorithm}[t]
		\caption{Module A Training for Heterogeneity-Aware Condition Encoding}
		\label{alg:module_a}
		\footnotesize
		\Input{training time-vehicle samples $\{(t,i,\mathbf{s}_t^i,b_t^i)\}$, encoder parameters $\phi_A$}
		\Output{updated encoder parameters $\phi_A$}
		\BlankLine
		\Repeat{the module-level stopping criterion is met}{
			sample a mini-batch of time-vehicle samples $(t,i)$\;
			\ForEach{time-vehicle sample $(t,i)$ in the mini-batch}{
				compute $\mathbf{c}_t^i=g_{\phi_A}(\mathbf{s}_t^i)$ from the four state branches\;
				form $\mathcal{S}_t^{i,+}$ and $\mathcal{S}_t^{i,-}$ according to $b_t^i$\;
				compute the stability and separation losses for this query sample\;
			}
			aggregate $\mathcal{L}^{\mathsf{repr}}$ and update $\phi_A$ by backpropagation\;
		}
	\end{algorithm}
	
	\vspace{-3mm}
	\subsection{Module B: Conditional Diffusion Trajectory Generation}\label{sec:module_b}
	
	Module B solves $\mathcal{P}_2$ by generating $K$ diverse executable candidates for each RI-vehicle pair. To this end, following topology-constrained diffusion generation \cite{controltraj2024} and diffusion-based closed-loop planning \cite{zheng2025fg}, DRIFT applies diffusion to the candidate control sequence rather than to the trajectory, so that the generated samples can be directly rolled out, feasibility-checked, and executed within the receding-horizon framework. Specifically, for the $k$-th candidate, we form the initial sample:
	\begin{equation}
		\mathbf{y}_0^{i,k}=\mathrm{vec}\!\left(\mathbf{U}_{t}^{i,k}\right),
	\end{equation}
	where $\mathbf{U}_{t}^{i,k}=\{\mathbf{u}_{t+h-1}^{i,k}\}_{h=1}^{H^{\mathsf{plan}}}$ is the candidate control sequence and $\mathrm{vec}(\cdot)$ stacks all control inputs into a single vector representation. With diffusion step $d=1,\dots,L^{\mathsf{diff}}$, the forward process follows the standard DDPM form \cite{ho2020ddpm,nichol2021improved}:
	\begin{equation}
		q\!\left(\mathbf{y}_{d}^{i,k}\mid \mathbf{y}_{d-1}^{i,k}\right)=
		\mathcal{N}\!\left(\sqrt{\alpha_{d}}\,\mathbf{y}_{d-1}^{i,k},\,\beta_{d}\mathbf{I}\right),
	\end{equation}
	meaning that the noisy sample at step $d$ is centered at the scaled previous sample $\sqrt{\alpha_d}\mathbf{y}_{d-1}^{i,k}$, while Gaussian noise with variance $\beta_d$ is injected into each dimension ($\alpha_d=1-\beta_d$). Hence, the \textit{forward diffusion process} gradually removes information from the original control sequence while accumulating noise, which can be represented with the following closed-form relationship:
	\begin{equation}
		\mathbf{y}_{d}^{i,k}=\sqrt{\bar{\alpha}_{d}}\,\mathbf{y}_0^{i,k}+\sqrt{1-\bar{\alpha}_{d}}\,\boldsymbol{\xi}^{i,k},
		\quad
		\bar{\alpha}_{d}=\prod_{r=1}^{d}\alpha_r,
	\end{equation}
	where $\boldsymbol{\xi}^{i,k}\sim\mathcal{N}(\mathbf{0},\mathbf{I})$. The \textit{reverse process} uses the condition representation learned by Module A to recover the original control sequence from the noisy sample. Specifically, a neural predictor parametrized by $\phi_B $  estimates the injected noise according to:
	\begin{equation}
		\widehat{\boldsymbol{\xi}}_{d}^{i,k}=\operatorname{Pred}_{\phi_B}\!\left(\mathbf{y}_{d}^{i,k},d,\mathbf{c}_t^{i}\right),
	\end{equation}
	from which the clean control and induced trajectory are reconstructed as:
	\begin{equation}
		\begin{aligned}
			\widehat{\mathbf{y}}_0^{i,k}&=\frac{1}{\sqrt{\bar{\alpha}_{d}}}
			\left(\mathbf{y}_{d}^{i,k}-\sqrt{1-\bar{\alpha}_{d}}\,\widehat{\boldsymbol{\xi}}_{d}^{i,k}\right),\\
			\widehat{\mathbf{U}}_{t}^{i,k}&=\mathrm{vec}^{-1}(\widehat{\mathbf{y}}_0^{i,k}),
		\end{aligned}
	\end{equation}
	where $\widehat{\mathbf{U}}_{t}^{i,k}=\{\widehat{\mathbf{u}}_{t+h-1}^{i,k}\}_{h=1}^{H^{\mathsf{plan}}}$. The corresponding trajectory is obtained by rolling the reconstructed controls forward through the transition model:
	\begin{equation}
		\widehat{\boldsymbol{\tau}}_{t}^{i,k}=
		\operatorname{Rollout}_{f}\!\left(\mathbf{x}_{t}^{i},\widehat{\mathbf{U}}_{t}^{i,k}\right).
	\end{equation}
	Thus, the diffusion model generates candidate control actions, while the resulting trajectories are obtained through the rollout operator $\operatorname{Rollout}_{f}$ in~\eqref{eq:R20} used by the closed-loop system in Section~\ref{sec:closed_loop_replanning}.\footnote{During training, the reference control sequence is taken from the control library or reconstructed from the expert/reference trajectory under the same transition model.}  Accordingly, the generator objective is given by:
	\begin{equation} \label{eq:LgenDef}
		\mathcal{L}^{\mathsf{gen}}(\phi_B)=\mathcal{L}^{\mathsf{noise}}(\phi_B)+\lambda^{\mathsf{fea}}\mathcal{L}^{\mathsf{fea}}(\phi_B),
	\end{equation}
	where 
	\begin{equation}\label{eq:LgenDef22}
		\hspace{-4mm}
		\resizebox{0.92\linewidth}{!}{$
			\mathcal{L}^{\mathsf{noise}}(\phi_B){=}
			\mathbb{E}_{i,t,k,d,\boldsymbol{\xi}}
			\left[w_t^{i,k}\left\|\boldsymbol{\xi}^{i,k}-
			\operatorname{Pred}_{\phi_B}(\mathbf{y}_{d}^{i,k},d,\mathbf{c}_t^{i})\right\|_2^2\right]$}\hspace{-4mm}
	\end{equation}
	is the standard diffusion denoising loss that trains the model to recover the injected noise and $w_t^{i,k}\ge 1$ denotes a candidate-specific training weight, which controls the contribution of candidate $k$ to the generator loss (it is initialized to one and later replaced by the Module-C long-tail coefficient $\chi_t^{i,k}$, which increases the influence of safety-critical candidates during training). In addition, $\mathcal{L}^{\mathsf{fea}}$ penalizes generated candidates whose rolled-out trajectories violate the executable-candidate requirements defined in Section~\ref{sec:problem_formulation}. All feasibility terms below are evaluated on the reconstructed pair $(\widehat{\mathbf{U}}_{t}^{i,k},\widehat{\boldsymbol{\tau}}_{t}^{i,k})$ produced by $\operatorname{Pred}_{\phi_B}$ and the rollout operator in \eqref{eq:R20}, whose one-step transition is defined in \eqref{eq:rollout_operator}; hence $\mathcal{L}^{\mathsf{fea}}$ depends on $\phi_B$ through both the generated controls and their induced rollout. To obtain $\mathcal{L}^{\mathsf{fea}}$, we first define the topological deviation of a rolled-out candidate as:
	\begin{equation}
		\Delta^{\mathsf{topo}}(\widehat{\boldsymbol{\tau}}_{t}^{i,k})
		=\frac{1}{H^{\mathsf{plan}}}\sum_{h=1}^{H^{\mathsf{plan}}}
		\mathbb{I}\!\left(\widehat{\mathbf{x}}_{t+h}^{i,k}\notin\Omega_{t+h}^{\mathsf{map}}\right),
		\label{eq:topo_deviation}
	\end{equation}
	where $\Omega_{t+h}^{\mathsf{map}}$ denotes the map-admissible vehicle-state set at TS $t+h$ induced by $\mathcal{G}^{\mathsf{map}}$, including road-boundary consistency, reachable lane membership, lane-connectivity constraints, and conflict-edge occupancy consistency.  $\Delta^{\mathsf{topo}}(\widehat{\boldsymbol{\tau}}_{t}^{i,k})$ is a differentiable-training counterpart of the lane-topology violation rate $r_t^{i,k,\mathsf{map}}$ in~\eqref{eq:main_map_violation}: it increases when the rollout generated by $\operatorname{Pred}_{\phi_B}$ leaves the map-admissible set or violates lane-topology consistency.
	We further define the boundary loss as follows:
	\begin{equation}
		\hspace{-4mm}\begin{aligned}
			&\ell_{t+h}^{i,k,\mathsf{bd}}
			(\widehat{\mathbf{U}}_{t}^{i,k},\widehat{\boldsymbol{\tau}}_{t}^{i,k})\\
			={}&
			\|[\widehat{\mathbf{u}}_{t+h-1}^{i,k}-\mathbf{u}^{\mathsf{max}}]_+\|_1
			+\|[\mathbf{u}^{\mathsf{min}}-\widehat{\mathbf{u}}_{t+h-1}^{i,k}]_+\|_1\\
			&+[\widehat v_{t+h}^{i,k}-v^{\mathsf{max}}]_+
			+[v^{\mathsf{min}}-\widehat v_{t+h}^{i,k}]_+,
		\end{aligned}
		\hspace{-4mm}
	\end{equation}
	where $\widehat v_{t+h}^{i,k}$ is the speed component of the predicted state $\widehat{\mathbf{x}}_{t+h}^{i,k}$ obtained from $\widehat{\boldsymbol{\tau}}_{t}^{i,k}$, and the $\ell_1$ norm sums the component-wise violations of the generated control bounds. Further, we define the safety-margin loss as follows:
	\begin{equation}
		\hspace{-4mm} \resizebox{0.93\linewidth}{!}{$
			\ell_{t}^{i,k,\mathsf{safe}}(\widehat{\boldsymbol{\tau}}_{t}^{i,k})=
			[\delta^{\mathsf{THW}}-\mathrm{THW}_{t}^{i,k,\mathsf{min}}]_{_+}
			+[\delta^{\mathsf{TTC}}-\mathrm{TTC}_{t}^{i,k,\mathsf{min}}]_{_+},
			$}\hspace{-4mm}
	\end{equation}
	which captures the THW and TTC safety-margin violation penalties computed along $\widehat{\boldsymbol{\tau}}_{t}^{i,k}$. When a diagnostic such as $r_t^{i,k,\mathsf{col}}(\widehat{\boldsymbol{\tau}}_{t}^{i,k})$ is written with an argument, the argument specifies the rollout on which the already-defined clearance diagnostic in \eqref{eq:main_collision_penalty} is evaluated. Finally, based on the above quantities, we obtain the feasibility objective $\mathcal{L}^{\mathsf{fea}}$ as:
	\begin{equation}
		\begin{aligned}
			&\mathcal{L}^{\mathsf{fea}}(\phi_B)={}
			\mathbb{E}_{i,t,k}\Bigg[
			w_t^{i,k}\Bigg(
			\frac{1}{H^{\mathsf{plan}}}\sum_{h=1}^{H^{\mathsf{plan}}}
			\ell_{t+h}^{i,k,\mathsf{bd}}
			(\widehat{\mathbf{U}}_{t}^{i,k},\widehat{\boldsymbol{\tau}}_{t}^{i,k})\\
			&+\lambda^{\mathsf{topo}}\,
			\Delta^{\mathsf{topo}}(\widehat{\boldsymbol{\tau}}_{t}^{i,k})
			+\lambda^{\mathsf{col}} r_t^{i,k,\mathsf{col}}(\widehat{\boldsymbol{\tau}}_{t}^{i,k})\\
			&+\lambda^{\mathsf{safe}}\ell_{t}^{i,k,\mathsf{safe}}(\widehat{\boldsymbol{\tau}}_{t}^{i,k})
			\Bigg)\Bigg].
		\end{aligned}
	\end{equation}
	Consequently, \eqref{eq:LgenDef} trains Module B to generate control sequences that are both diffusion-consistent and executable, while encouraging their rolled-out trajectories to satisfy actuation limits, speed limits, roadway-topology constraints, and THW/TTC safety requirements. Algorithm~\ref{alg:module_b} summarizes this procedure. During online rollout, $K$ candidates are sampled, rolled out, filtered by $\Omega_t^{i,\mathsf{feas}}$, scored according to \eqref{eq:score_j}, and selected through $\mathcal{P}_{\mathsf{sel}}$. 
	
	\begin{algorithm}[t]
		\caption{Module B Training and Candidate Generation}
		\label{alg:module_b}
		\footnotesize
		\Input{time-vehicle sample batch $\{(t,i,\mathbf{s}_t^i,\mathbf{x}_t^i)\}$, reference controls or trajectories, cached weights $\{w_t^{i,k}\}$, generator parameters $\phi_B$, frozen encoder $\phi_A$}
		\Output{updated generator parameters $\phi_B$}
		\BlankLine
		\Repeat{the module-level stopping criterion is met}{
			sample a time-vehicle mini-batch\;
			\ForEach{time-vehicle sample $(t,i)$ in the mini-batch}{
				compute the condition vector $\mathbf{c}_t^i=g_{\phi_A}(\mathbf{s}_t^i)$\;
				\For{$k\leftarrow 1$ \KwTo $K$}{
					form the clean/reference control vector $\mathbf{y}_{0}^{i,k}=\mathrm{vec}(\mathbf{U}_{t}^{i,k})$\;
					sample $(d,\boldsymbol{\xi}^{i,k})$, construct $\mathbf{y}_{d}^{i,k}$, and predict $\widehat{\boldsymbol{\xi}}_{d}^{i,k}$\;
					reconstruct $\widehat{\mathbf{U}}_{t}^{i,k}$ and roll it out into $\widehat{\boldsymbol{\tau}}_{t}^{i,k}$\;
					accumulate the weighted denoising and feasibility losses\;
				}
			}
			form $\mathcal{L}^{\mathsf{gen}}=\mathcal{L}^{\mathsf{noise}}+\lambda^{\mathsf{fea}}\mathcal{L}^{\mathsf{fea}}$\;
			update $\phi_B$ by backpropagation\;
		}
	\end{algorithm}
	
	\subsection{Module C: Progressive Adversarial Imitation with Long-Tail-Risk Enhancement}\label{sec:module_c}
	
	Module C addresses $\mathcal{P}_3$ by supplementing the candidate generator with distribution alignment and long-tail-risk feedback. Its goal is to encourage generated candidates to remain consistent with expert driving behaviors while increasing attention to rare safety-critical situations. During training, Module C learns a state-conditioned discriminator that distinguishes expert trajectories from generated trajectories under the same traffic context. During inference, the trained discriminator is reused as the behavior-similarity scorer in \eqref{eq:score_j}. Given the fixed generator $\phi_B^{\mathsf{fix}}$ from Module B, Module C compares generated candidate trajectories against expert driving trajectories and gradually introduces an adversarial alignment objective during training. Specifically, 
	we first define the base alignment loss as follows:
	\begin{equation} \label{eq:BaseAlignDef}
		\begin{aligned}
			\mathcal{L}^{\mathsf{align}}(\vartheta;\phi_B^{\mathsf{fix}})={}&
			\mathbb{E}_{(t,i)}\Bigg[
			-\log\!\Big(\sigma\!\big(\mathcal{D}_{\vartheta}(\boldsymbol{\tau}_{t}^{i,\mathrm{exp}},\mathbf{s}_t^{i})\big)\Big)\\
			&\quad-\frac{1}{K}\sum_{k=1}^{K}
			\log\!\Big(1-\sigma\!\big(\mathcal{D}_{\vartheta}(\widehat{\boldsymbol{\tau}}_{t}^{i,k},\mathbf{s}_t^{i})\big)\Big)
			\Bigg],
		\end{aligned}
	\end{equation}
	where $\boldsymbol{\tau}_{t}^{i,\mathrm{exp}}$ denotes the expert/reference trajectory and $\widehat{\boldsymbol{\tau}}_{t}^{i,k}$ denotes the $k$-th generated trajectory produced by the fixed generator $\phi_B^{\mathsf{fix}}$. In \eqref{eq:BaseAlignDef},    
	$\mathcal{D}_{\vartheta}(\boldsymbol{\tau},\mathbf{s})$ is a discriminator network that outputs a \textit{realism score} indicating how closely trajectory $\boldsymbol{\tau}$ resembles expert driving behavior under state $\mathbf{s}$, and $\sigma(\cdot)$ is the Sigmoid function.     
	To gradually introduce the adversarial objective during training, we define:
	\begin{equation}\label{eq:adv_warmup}
		\lambda^{\mathsf{adv},(e)}=\min\!\left(1,\frac{e}{E^{\mathsf{warm}}}\right),
	\end{equation}
	where $e$ denotes the current Module-C training epoch and $E^{\mathsf{warm}}$ denotes the number of warm-up epochs over which the adversarial alignment term is increased from zero to its full weight. 
	Finally, Module C is trained according to the objective:
	\begin{equation}\label{eq:LAE}
		\mathcal{L}^{\mathsf{align},(e)}(\vartheta;\phi_B^{\mathsf{fix}})=\lambda^{\mathsf{adv},(e)}\mathcal{L}^{\mathsf{align}}(\vartheta;\phi_B^{\mathsf{fix}}).
	\end{equation}
	Here, $\phi_C$ is used as a compact notation for the Module-C training block. Its trainable parameter block is the discriminator parameter $\vartheta$ in $\mathcal{D}_{\vartheta}$. The non-trainable Module-C quantities maintained during training are the warm-up variables $\lambda^{\mathsf{adv},(e)}$, $e$, and $E^{\mathsf{warm}}$ in~\eqref{eq:adv_warmup}, together with the cached long-tail coefficients $\{\chi_t^{i,k}\}_{k=1}^{K}$ defined in~\eqref{eq:chi_tail} and later copied into the Module-B weights as $w_t^{i,k}\leftarrow\chi_t^{i,k}$. No additional Module-C parameters are implied by this notation. 
	% The warm-up schedule defined by $\lambda^{\mathsf{adv},(e)}$ and $E^{\mathsf{warm}}$ only describes the progressive activation of the adversarial loss.
	The above adversarial formulation follows imitation-learning principles \cite{gail2016} and is conceptually related to diffusion-enhanced AIL \cite{diffail2024}. However, unlike conventional adversarial imitation learning, the discriminator output is ultimately used to evaluate executable candidate trajectories during candidate selection.

	To emphasize rare but safety-critical traffic situations and connect candidate-level failures with the rollout-level tail-risk penalty in \eqref{eq:p0tail}, Module C first computes a deterministic risk-aware coefficient:
	\begin{equation}\label{eq:chi_tail}
		\chi_t^{i,k}=1+\beta^{\mathsf{lt}}\big(R_t^{i,k}+D_t^{i,k}\big),
	\end{equation}
	where $R_t^{i,k}$ and $D_t^{i,k}$ were introduced in~\eqref{eq:score_j}, and $\beta^{\mathsf{lt}}\ge 0$ is a long-tail reweighting gain controlling the effect of risk and execution difficulty. The coefficient $\chi_t^{i,k}$ is not a learnable parameter; it is computed from the measured $R_t^{i,k}$ and $D_t^{i,k}$ for the current generated candidate. Larger values of $\chi_t^{i,k}$ assign greater importance to candidates associated with elevated safety risk or operational difficulty. Using this deterministic coefficient, the corresponding long-tail weighted discrepancy is:
	\begin{equation}\label{eq:Ltail}
		\mathcal{L}^{\mathsf{tail}}=
		\mathbb{E}_{(t,i)}\!\left[
		\frac{1}{K}\sum_{k=1}^{K}
		\chi_t^{i,k}\,
		\left\|\widehat{\boldsymbol{\tau}}_{t}^{i,k}-\boldsymbol{\tau}_{t}^{i,\mathrm{exp}}\right\|_1
		\right].
	\end{equation}
	We note that the objective in~\eqref{eq:p3} is used to update the discriminator parameters $\vartheta$ through $\mathcal{L}^{\mathsf{align},(e)}(\vartheta;\phi_B^{\mathsf{fix}})$ in \eqref{eq:LAE}, while $\mathcal{L}^{\mathsf{tail}}$ in \eqref{eq:Ltail} is not a separate optimizer; instead, it defines the finite-sample long-tail signal used by the stagewise training loop: after Module C evaluates a mini-batch, the deterministic coefficients $\{\chi_t^{i,k}\}_{k=1}^{K}$ are cached and then substituted into the next Module-B update as $w_t^{i,k}\leftarrow\chi_t^{i,k}$ in~\eqref{eq:LgenDef22} (i.e., they determine which generated candidates receive larger weights in the following generator update). 
	% Therefore, Eq.~\eqref{eq:p3} trains the discriminator, while $\mathcal{L}^{\mathsf{tail}}$ determines which generated candidates receive larger weights in the following generator update. 
	This is the mechanism by which the tail-risk target in~\eqref{eq:p0tail} affects candidate generation. Algorithm~\ref{alg:module_c} summarizes this procedure. During online rollout, Module C performs no adversarial updates; the discriminator output $\mathcal{D}_{\vartheta}$ contributes to behavior similarity, while $R_t^{i,k}$ and $D_t^{i,k}$ remain explicit candidate-selection terms in $\mathcal{P}_{\mathsf{sel}}$.
	
	\begin{algorithm}[t]
		\caption{Module C Stagewise Alignment and Long-Tail Reweighting}
		\label{alg:module_c}
		\footnotesize
		\Input{expert trajectories $\{\boldsymbol{\tau}_{t}^{i,\mathrm{exp}}\}$, generated candidates $\{\widehat{\boldsymbol{\tau}}_{t}^{i,k}\}$, local states $\{\mathbf{s}_t^i\}$, fixed generator $\phi_B$, Module-C discriminator parameters $\vartheta$}
		\Output{updated discriminator parameters $\vartheta$ and cached reweighting coefficients $\chi_t^{i,k}$}
		\BlankLine
		\Repeat{the module-level stopping criterion is met}{
			advance the epoch counter $e$ and set $\lambda^{\mathsf{adv},(e)}=\min(1,e/E^{\mathsf{warm}})$\;
			\ForEach{mini-batch of time-vehicle samples $(t,i)$}{
				fetch expert and generated trajectories with local state $\mathbf{s}_t^i$\;
				\For{$k\leftarrow 1$ \KwTo $K$}{
					evaluate discriminator realism and candidate risk/violation terms\;
					compute $\chi_t^{i,k}=1+\beta^{\mathsf{lt}}(R_t^{i,k}+D_t^{i,k})$\;
				}
				update the discriminator parameters $\vartheta$ with $\mathcal{L}^{\mathsf{align},(e)}$ while keeping $\phi_B$ fixed\;
				form the long-tail weighted discrepancy $\mathcal{L}^{\mathsf{tail}}$ and cache $\{\chi_t^{i,k}\}_{k=1}^{K}$ for the next Module-B update\;
			}
		}
	\end{algorithm}
	
	\section{Experiments}\label{sec:experiments}
	
	\newcommand{\experimentsettingitem}[1]{%
		\par\noindent\hangindent=\parindent\hangafter=1%
		\makebox[\parindent][l]{$\bullet$}\textbf{#1}: }
	\newcommand{\experimentmethoditem}[2]{%
		\par\noindent\hangindent=\parindent\hangafter=1%
		\makebox[\parindent][l]{$\bullet$}\textit{#1}: #2\par}
	\newcommand{\resultfigwidth}{\columnwidth}
	
	\subsection{Experimental Setup}\label{subsec:exp_setup}
	
	We evaluate DRIFT in closed-loop mixed-autonomy simulation,
	% focusing on three aspects: (i) efficiency across AV penetration rates, (ii) safety under merging and strong vehicle-interaction disturbances, and (iii) the contribution of each module to traffic-level outcomes. 
	% All methods are executed in the same \textit{Flow}/\textit{SUMO} feedback loop \cite{flow2022}, and performance is assessed by induced traffic evolution under executable controls. 
	% All experiments 
	following the settings below.
	
	\experimentsettingitem{Experimental platform} All experiments are implemented in \textit{Flow}/\textit{SUMO} \cite{flow2022}. At each RI, the controller observes the current traffic state and outputs the next executable action segment. After execution, the environment immediately enters the next HV/AV interaction round, so the evaluation preserves feedback effects such as hard braking, local congestion, disturbance propagation, and merge conflicts.
	
	\experimentsettingitem{Traffic scenarios} We consider three \textit{Flow}/\textit{SUMO} mixed-autonomy benchmark traffic scenarios. (i) \textit{Ring} follows the closed-road setting widely used for traffic-wave stabilization with AVs \cite{stern2018field} and tests disturbance propagation and stable car-following. (ii) \textit{Figure-eight} (F8) is a closed-network \textit{Flow} benchmark that captures strong local vehicle interactions. (iii) \textit{Merge} is an open-network highway-merge benchmark that evaluates yield coordination, conflict resolution, and traffic outflow. Collectively, these scenarios capture closed-network disturbance propagation, closed-network local conflicts, and open-network merging; evaluating larger urban networks is beyond the scope of this paper.
	
	\begin{figure}[!t]
		\vspace{-4mm}
		\centering
		\includegraphics[width=\resultfigwidth]{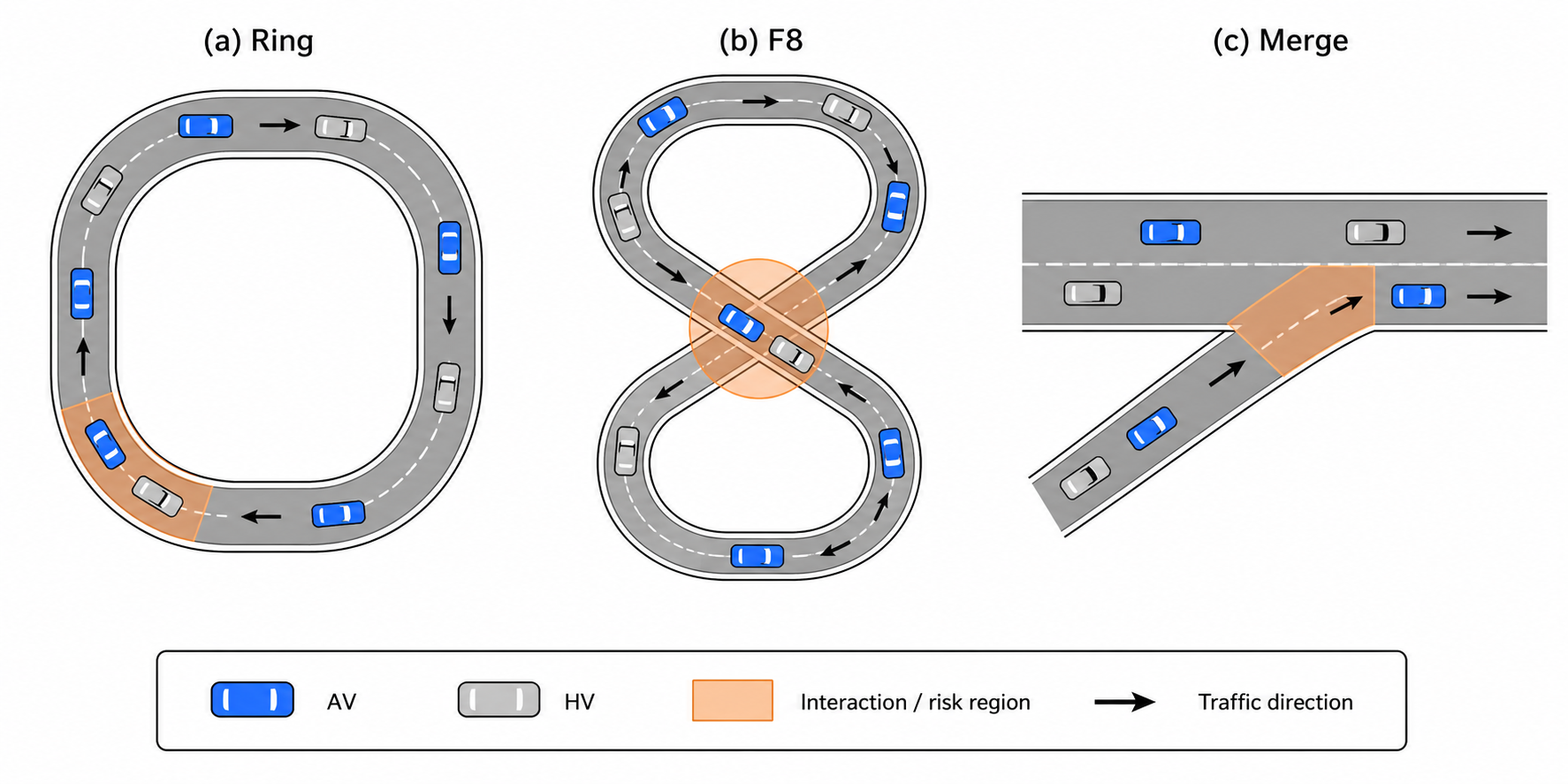}
		\vspace{-9mm}
		\caption{Representative benchmark scenarios. Blue vehicles denote AVs, gray vehicles denote HVs, and orange regions indicate disturbance, interaction, or merge-conflict areas.}
		\vspace{-5mm}
		\label{fig:experiment_scenarios}
	\end{figure}
	
	\experimentsettingitem{Real-trajectory datasets} The HV behavioral prior is calibrated from highD \cite{highd2018}, rounD \cite{round2020}, exiD \cite{exid2022}, and inD \cite{ind2020}. These datasets cover highway, roundabout, highway-ramp, and urban-intersection driving segments, and are used to estimate HV kinematic distributions $\mathbf{z}^{\mathsf{HV}}_s$ in~\eqref{eq:hv_calibration}, local-interaction features $\mathbf{n}_t^i$ in~\eqref{eq:NII}, and safety/risk quantities such as THW, TTC, $R_t^{i,k}$, and $D_t^{i,k}$ in~\eqref{eq:score_j}.
	
	\experimentsettingitem{AV penetration} Closed-loop evaluation is conducted under six AV penetration rates, namely $0\%, 20\%, 40\%, 60\%, 80\%,$ and $100\%$. Each scenario-method-penetration is repeated for five closed-loop episodes. 
	% Flow-RL is not applicable at $0\%$ penetration because it requires at least one RL-controlled vehicle.
	
	\experimentsettingitem{HV behavior extraction and calibration} Rather than replaying recorded trajectories frame by frame, we extract vehicle class, speed, longitudinal acceleration, longitudinal/lateral velocity, THW, TTC, and lane-change indicators from the raw tracks and metadata files. For each scenario $s$, the calibrated HV prior is represented as:
	\begin{equation}
		\begin{aligned}
			\mathbf{z}^{\mathsf{HV}}_s
			=[\hat v_s^{\mathsf{des}},\hat T_s,\hat s_{0,s},
			\hat a_s^{\mathsf{max}},\hat b_s,\hat\tau_s,\hat\sigma_s],
		\end{aligned}
		\label{eq:hv_calibration}
	\end{equation}
	where $\hat{\cdot}$ is statistics calibrated from real trajectories; $v^{\mathsf{des}}$, $T$, $s_0$, $a^{\mathsf{max}}$, $b$, $\tau$, and $\sigma$ denote desired speed, time headway, minimum spacing, maximum acceleration, comfortable deceleration, reaction delay, and control noise, respectively. Training samples use a history window of length 6 and a future window of length 6 with stride 5. 
	% Vehicle type and AV penetration correspond to the control-authority marker and penetration embedding in the formal state definition, while scenario differences are mainly represented by road semantics; an additional scenario identifier is used only for sample construction, bucketing, and experimental logging.
	
	\experimentsettingitem{AV behavior generation} AV behavior is not directly sampled as a native category from the real trajectory datasets. Instead, parameterized AV types are generated from the calibrated HV prior with mixed-autonomy and scenario-specific adjustments:
	\begin{equation}
		\begin{aligned}
			\mathbf{z}^{\mathsf{AV}}_s
			=\big[&\min(v_s^{\mathsf{lim}},\alpha_s^{\mathsf{v}}\hat v_s^{\mathsf{des}}),
			\alpha_s^{\mathsf{T}}\hat T_s,\alpha_s^{\mathsf{s0}}\hat s_{0,s},\\
			&\alpha_s^{\mathsf{a}}\hat a_s^{\mathsf{max}},\alpha_s^{\mathsf{b}}\hat b_s,
			\alpha_s^{\mathsf{tau}}\hat\tau_s,\alpha_s^{\mathsf{sigma}}\hat\sigma_s\big],
		\end{aligned}
		\label{eq:av_generation}
	\end{equation}
	where the $\alpha$ coefficients denote scenario-dependent scaling and $v_s^{\mathsf{lim}}$ is the road speed limit. Thus, $\mathbf{z}^{\mathsf{AV}}_s$ contains the same seven behavior-prior entries as $\mathbf{z}^{\mathsf{HV}}_s$: desired speed, time headway, minimum spacing, maximum acceleration, comfortable deceleration, reaction delay, and control noise. During online closed-loop execution, HVs follow the calibrated behavioral prior, while controlled AVs read a seven-dimensional history feature from \textit{Flow}/\textit{SUMO}: speed, estimated acceleration, leader headway, leader-relative speed, THW, TTC, and lane-change state. This history feature is used by DRIFT to generate candidates, construct the feasible candidate set, score and select candidates, and execute the selected control segment.
	
	% The AV prior is therefore a bounded synthetic prior for penetration-rate sensitivity rather than an empirical AV-fleet model. 
	
\begin{figure}[!t]
		\vspace{-4mm}
		\centering
		\includegraphics[width=0.94\columnwidth]{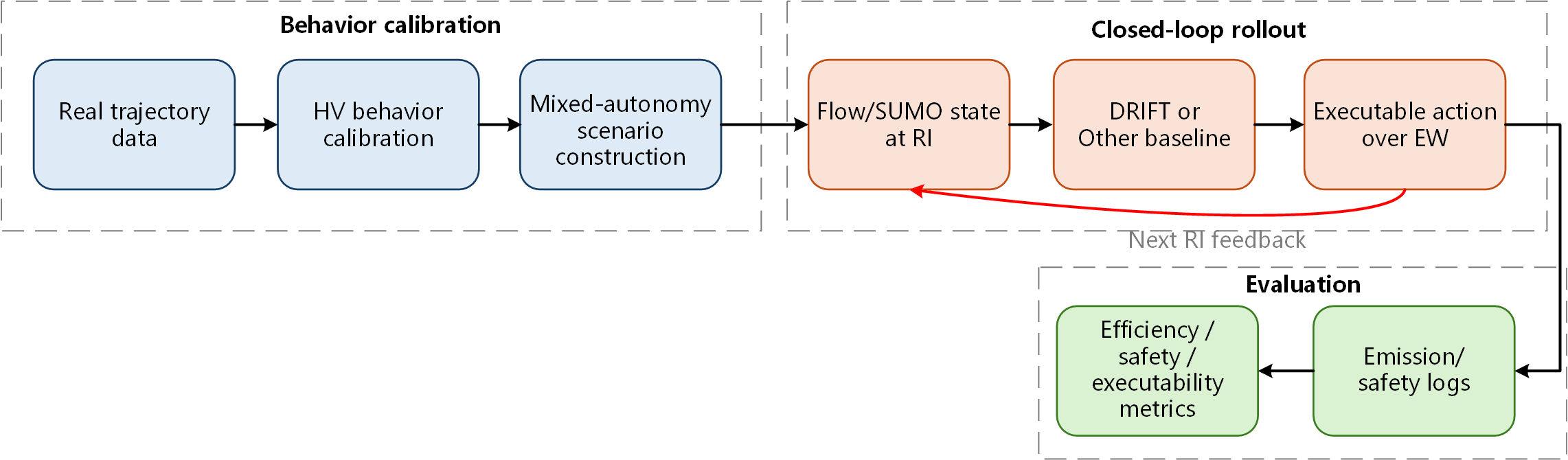}
		\vspace{-3mm}
		\caption{Closed-loop simulation protocol. Real trajectory data calibrates HV behavior and AV priors; at each RI, DRIFT or a baseline reads the \textit{Flow}/\textit{SUMO} state, outputs executable actions, and logs efficiency, safety, and executability metrics.}
		\label{fig:closed_loop_protocol}
		\vspace{-5mm}
	\end{figure}

	The scaling ratios in Table~\ref{tab:av_prior_scaling} of Appendix~\ref{sec:appendix_implementation} show that the desired speed gain is mild, reaction delay is reduced but nonzero, and control noise remains positive. This keeps the \textit{Flow}/\textit{SUMO} experiments reproducible. 
	% Overall, this setup ensures that real trajectory data are used for behavior calibration and model training, while the main results are obtained from traffic evolution induced by executable closed-loop controls under each penetration setting. 
	Detailed protocol, metric, and reproducibility settings are reported in Appendix~\ref{sec:appendix_implementation} through Tables~\ref{tab:experiment_protocol} and~\ref{tab:reproducibility_protocol} and an overview of our implementation pipeline is depicted in Fig.~\ref{fig:closed_loop_protocol}. In the final implementation, output safety filtering in high-penetration \textit{F8} cases is treated as a scenario-specific execution boundary of the feasible set $\Omega_t^{i,\mathsf{feas}}$, while ETA-yield coordination in \textit{Merge} is treated as a scenario-specific realization of the merge-conflict-related dynamic/topological cost $D_t^{i,k}$ and the online selection problem $\mathcal{P}_{\mathsf{sel}}$. They are not additional modules outside DRIFT and are not reported as separate methods.
	
	\subsection{Baselines and Variants}\label{subsec:baselines_variants}
	
	We compare three groups of methods. The first group contains external baselines that can be directly executed in the \textit{Flow}/\textit{SUMO} closed-loop. The second group contains DRIFT and simple internal replacements. The third group contains ablation variants used to explain module and mechanism contributions. These groups are detailed below.
	
	\subsubsection{External methods and baselines} We consider the following baseline methods in our analyses:
	\experimentmethoditem{FollowerStopper}{A classical rule-based controller for mixed-autonomy ring-road stabilization \cite{stern2018field}, used as a \textit{Flow}-native closed-loop baseline.}
	\experimentmethoditem{PI controller}{The \texttt{PISaturation} controller provided by \textit{Flow} \cite{flow2022}, used as a simple feedback-control baseline.}
	% It is an executable engineering controller rather than a newly proposed standalone paper method.}
\experimentmethoditem{IDM}{The Intelligent Driver Model car-following controller in \textit{Flow} \cite{flow2022}, included as a classical car-following reference under the same executable feedback interface.}
\experimentmethoditem{Flow-RL}{A learning-control baseline constructed from the official reinforcement learning (RL)-controlled vehicle interface in \textit{Flow} \cite{flow2022}. Since it requires at least one RL-controlled vehicle, the $0\%$ AV-penetration cell is not applicable.}
\experimentmethoditem{Flow-AIL}{An imitation-style learning baseline adapted to the same \textit{Flow}/\textit{SUMO} executable interface. It uses the same observation variables as DRIFT, learns a direct state-to-control mapping from the calibrated expert/control traces with an adversarial imitation loss, and outputs a single executable control segment at each RI rather than a feasibility-filtered candidate set.}

These baselines are selected according to closed-loop executability and protocol consistency. The comparisons therefore include methods that can all read the same \textit{Flow}/\textit{SUMO} traffic state and return executable controls at every RI. 
% Broader traffic-generation methods discussed in Section~\ref{sec:core_motivation} are not inserted into this numerical table because they use different interfaces, such as offline scene generation, initialization synthesis, or search-based scenario construction.}

\subsubsection{Internal variants} We consider the full implementation of DRIFT and a set of variations where its components/modules are replaced with simpler designs as follows:
\experimentmethoditem{DRIFT}{The full proposed method with Module A, Module B, and Module C.}
\experimentmethoditem{Simple-A}{A variant that replaces Module A with a plain MLP encoder using only ego-history and current kinematic variables, without the heterogeneous interaction, road-semantic, and penetration embeddings in~\eqref{eq:HLS}, while retaining learned Modules B/C, i.e., BC+simple A.}
\experimentmethoditem{Simple-B}{A variant that replaces Module B with a fixed candidate generator that samples acceleration/yield candidates from the calibrated control prior instead of using the conditional diffusion generator, while retaining learned Modules A/C, i.e., AC+simple B.}
\experimentmethoditem{Simple-C}{A variant that replaces Module C with a heuristic candidate-selection, i.e., selecting candidates only by efficiency and rule-based safety penalties without discriminator-based behavior similarity or long-tail reweighting, while retaining learned Modules A/B, i.e., AB+simple C.}

\subsubsection{Ablation settings} We consider the following ablation settings where various components/modules of DRIFT are excluded from the design:
\experimentmethoditem{w/o Module A}{Removes the condition-encoding module from DRIFT and keeps only the downstream generation and scoring interfaces needed for closed-loop execution.}
\experimentmethoditem{w/o Module B}{Removes the learned candidate-generation backbone from DRIFT and uses the necessary execution interface to maintain closed-loop rollout.}
\experimentmethoditem{w/o Module C}{Removes the discriminator-based realism score and long-tail feedback provided by Module C from DRIFT, while retaining the basic risk and dynamic-cost terms required by $\mathcal{P}_{\mathsf{sel}}$ for closed-loop execution.}
\experimentmethoditem{w/o Penetration Conditioning}{Removes AV-penetration conditioning from DRIFT to test cross-penetration adaptation.}
\experimentmethoditem{Single-Candidate}{Degenerates multi-candidate generation of DRIFT into single-candidate execution to test the value of candidate diversity and selection.}

We note that the aforementioned internal replacements and ablations have different meanings: the former replaces one module with a simple counterpart to test whether the learned design is better than a simple substitute, whereas the latter removes the corresponding module  to test whether that module is necessary. 
% Specifically, Simple-A uses a plain non-heterogeneity-aware condition encoder, Simple-B uses a fixed candidate generator rather than learned diffusion sampling, and Simple-C uses a heuristic selection surrogate rather than the learned Module-C realism/risk-feedback interface. By contrast, w/o Module A/B/C removes the corresponding learned module position while retaining only the minimal execution interface needed to keep the simulator rollout comparable. 
Appendix~\ref{sec:appendix_implementation} summarizes the role, coverage, and implementation status of all compared methods in Table~\ref{tab:method_status}.

% To keep the experimental section consistent with the executable controller, the reported Module B uses the executable-control parameterization described in Section~\ref{sec:module_b}: candidate controls are generated in the compact control space, rolled out into finite-horizon trajectory candidates, and then passed to feasibility filtering, risk/dynamic-cost evaluation, Module-C realism scoring, and $\mathcal{P}_{\mathsf{sel}}$. Thus, all reported results correspond to executable controls under the same closed-loop interface rather than to offline trajectory samples evaluated only by open-loop similarity.

\subsection{Evaluation Metrics and Analysis}\label{subsec:metrics}

\subsubsection{Metric Definitions}

% The evaluation metrics cover efficiency, safety, closed-loop executability, and stability diagnostics. 
Efficiency metrics include \textit{episode return} (the accumulated rollout score associated with the selected-candidate objective in~\eqref{eq:p0}), \textit{average speed} (measures traffic-flow efficiency),  and \textit{outflow} (measures throughput in the open merge network). Safety and executability diagnostics (computed from the corresponding emission and safety logs) include \textit{collision/deadlock, hard braking, minimum realized acceleration, THW,} and \textit{TTC}. 
% Closed-loop stability is assessed through penetration-wise smoothness of return/speed/outflow trends together with the episode-level variability reported in Appendix~\ref{sec:appendix_supp_results}. 
% Each table cell is aggregated from five closed-loop episodes unless otherwise stated, and safety diagnostics are computed from the corresponding emission and safety logs. All tables mark the preferred direction of each metric and bold the best available value under that direction; 
Entries marked by ``--'' in the tables or ``Not applicable" in the figures indicate that the metric is not exposed by the corresponding evaluation interface or is not applicable to that scenario.

Efficiency results are interpreted together with safety diagnostics because high short-term speed or outflow may be accompanied by hard braking, extreme deceleration, or simulator safe-speed clipping. We therefore draw conclusions from the joint pattern of return, speed/outflow, TTC risk, hard braking, and minimum acceleration rather than from a single efficiency metric. Additional statistical, sensitivity, latency, selector/feasibility, compact stress-test, and validation-set diagnostics are reported in Appendix~\ref{sec:appendix_supp_results}.

The main comparisons are reported in Table~\ref{tab:main_summary}. Figs.~\ref{fig:efficiency_trends_summary}--\ref{fig:ablation_delta_summary} then separate the same evidence into penetration-wise efficiency trends, braking and acceleration diagnostics, and module-level ablation effects.

\subsubsection{Performance Overview}\label{subsec:main_results}

\begin{figure}[!t]
% \vspace{-8mm}
\centering
\includegraphics[width=\columnwidth]{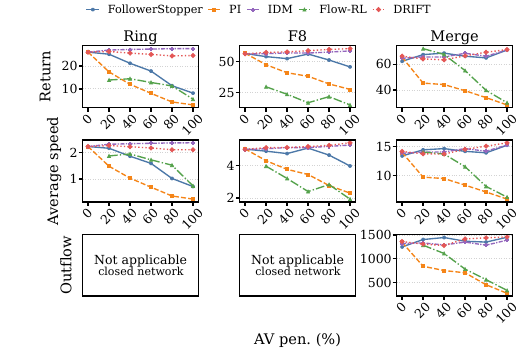}
\vspace{-8mm}
\caption{Efficiency trends across various AV penetration rates captured by the x-axis of the plots. Rows report return, average speed, and \textit{Merge} outflow; columns denote traffic scenarios. Outflow is reported only for \textit{Merge}.}
\label{fig:efficiency_trends_summary}
\vspace{-3mm}
\end{figure}

Table~\ref{tab:main_summary} shows that DRIFT is not uniformly best on every isolated metric, but it provides a favorable joint pattern across efficiency and safety/executability. In \textit{Ring}, IDM achieves the highest return and speed, while DRIFT remains close to IDM and keeps zero collision, zero hard-braking events, and the least severe Worst $a$. In \textit{F8}, Flow-AIL slightly leads in return and speed, but DRIFT reduces the close-interaction exposure reported by TTC\% and THW\% and also lowers HB. In \textit{Merge}, Flow-AIL reaches the largest outflow, but DRIFT is nearly tied in outflow while reducing HB from 78 to 26 and improving Worst $a$ from $-100.27$ to $-97.40$. Fig.~\ref{fig:efficiency_trends_summary} further shows that the return, speed, and outflow curves vary smoothly across AV penetration levels, which supports the role of heterogeneity-aware conditioning under changing mixed-autonomy compositions.

% Our main results are provided in 
% Table~\ref{tab:main_summary}, while Fig.~\ref{fig:efficiency_trends_summary} reports efficiency trends across six AV penetration rates. In Fig.~\ref{fig:efficiency_trends_summary}, the return and average speed rows show that DRIFT changes smoothly with AV penetration in \textit{Ring} and \textit{F8}, without abrupt penetration-wise jumps. Also, in \textit{Merge}, the outflow row of Fig.~\ref{fig:efficiency_trends_summary} should be interpreted together with the safety columns of Table~\ref{tab:main_summary}: Flow-AIL reaches the highest outflow, whereas DRIFT remains nearly tied in outflow while having fewer hard-braking events (HB) and a less severe worst realized acceleration (Worst $a$). The average speed and outflow plots in Fig.~\ref{fig:efficiency_trends_summary} further indicate closed-network behavior: IDM remains most efficient in \textit{Ring}, Flow-AIL slightly leads in \textit{F8} efficiency but has larger close-interaction percentages in Table~\ref{tab:main_summary} (TTC\% and THW\%), and FollowerStopper remains competitive in \textit{Merge}. Thus, DRIFT is not designed to maximize every single metric; rather, it maintains near-best efficiency while improving the safety/executability (reported by the columns in Table~\ref{tab:main_summary} that measure hard braking, close interaction, and extreme deceleration).

\begin{table}[!t]
\vspace{-4mm}
\centering
\caption{Main results with safety and executability diagnostics. Here Ret., Spd., and Out. denote episode return, average speed, and outflow; Coll. denotes collision/deadlock count; TTC\% and THW\% denote the percentages of time steps below the TTC and THW safety thresholds; HB denotes hard-braking count; Worst $a$ denotes the minimum realized acceleration.}
\label{tab:main_summary}
\vspace{-2mm}
\begingroup
\setlength{\tabcolsep}{1.0pt}
\renewcommand{\arraystretch}{0.92}
\scriptsize
\begin{tabular}{llrrrrrrrr}
\toprule
Scenario & Method & Ret.$\uparrow$ & Spd.$\uparrow$ & Out.$\uparrow$ & Coll.$\downarrow$ & TTC\%$\downarrow$ & THW\%$\downarrow$ & HB$\downarrow$ & Worst $a$ $\uparrow$ \\
\midrule
Ring & FollowerStopper & 18.32 & 1.59 & -- & \textbf{0} & 1.03 & \textbf{0.00} & 87 & -57.89 \\
Ring & PI & 11.81 & 1.01 & -- & \textbf{0} & \textbf{0.00} & \textbf{0.00} & \textbf{0} & -3.80 \\
Ring & IDM & \textbf{27.21} & \textbf{2.31} & -- & \textbf{0} & \textbf{0.00} & \textbf{0.00} & \textbf{0} & -0.92 \\
Ring & Flow-RL & 11.59 & 1.56 & -- & -- & 0.02 & \textbf{0.00} & \textbf{0} & -1.99 \\
Ring & Flow-AIL & 15.02 & 1.35 & -- & \textbf{0} & \textbf{0.00} & \textbf{0.00} & \textbf{0} & -4.80 \\
Ring & DRIFT & 25.41 & 2.17 & -- & \textbf{0} & \textbf{0.00} & \textbf{0.00} & \textbf{0} & \textbf{-0.91} \\
\midrule
F8 & FollowerStopper & 52.64 & 4.72 & -- & \textbf{0} & 0.11 & 0.22 & 52 & -102 \\
F8 & PI & 40.50 & 3.60 & -- & \textbf{0} & \textbf{0.03} & 0.10 & 17 & \textbf{-46.98} \\
F8 & IDM & 57.05 & 5.11 & -- & \textbf{0} & 0.17 & 0.18 & 16 & -78.39 \\
F8 & Flow-RL & 21.43 & 2.87 & -- & -- & 0.21 & \textbf{0.08} & \textbf{8} & -72.57 \\
F8 & Flow-AIL & \textbf{58.54} & \textbf{5.28} & -- & \textbf{0} & 1.05 & 0.85 & 30 & -72.43 \\
F8 & DRIFT & 58.38 & 5.17 & -- & \textbf{0} & 0.80 & 0.64 & 22 & -71.01 \\
\midrule
Merge & FollowerStopper & 66.73 & 14.31 & 1378 & \textbf{0} & 1.02 & 0.20 & 117 & -109 \\
Merge & PI & 42.85 & 9.14 & 726 & \textbf{0} & \textbf{0.08} & \textbf{0.06} & 162 & -106 \\
Merge & IDM & 66.91 & 14.36 & 1328 & \textbf{0} & 0.08 & 0.16 & 36 & \textbf{-97.40} \\
Merge & Flow-RL & 52.93 & 10.77 & 812 & -- & 0.80 & 0.12 & \textbf{19} & -102 \\
Merge & Flow-AIL & \textbf{67.38} & 14.48 & \textbf{1380} & \textbf{0} & 0.17 & 0.18 & 78 & -100.27 \\
Merge & DRIFT & 66.80 & \textbf{14.49} & 1377 & \textbf{0} & 0.16 & 0.10 & 26 & \textbf{-97.40} \\
\bottomrule
\end{tabular}

\endgroup
\vspace{-4mm}
\end{table}

\subsubsection{Safety and Executability Diagnostics}\label{subsec:safety_diagnostics}
\begin{figure}[!t]
% \vspace{-5mm}
\centering
\includegraphics[width=\columnwidth]{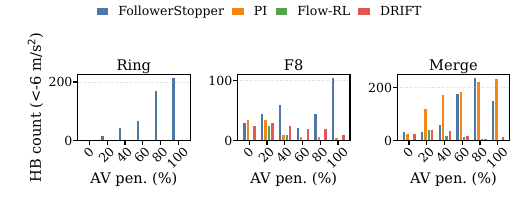}
\vspace{-10mm}
\caption{Hard-braking (HB) diagnostics. HB$<-6$ denotes the number of logged time steps whose realized longitudinal acceleration is below $-6\,\mathrm{m/s^2}$.}
\label{fig:hard_braking_bars}
\vspace{-4mm}
\end{figure}

\begin{figure}[!t]
% \vspace{-5mm}
\centering
\includegraphics[width=\columnwidth]{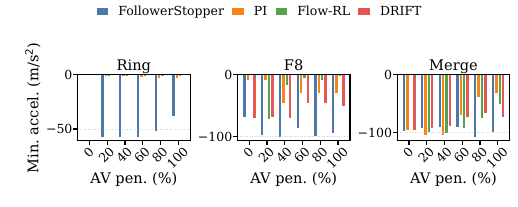}
\vspace{-10mm}
\caption{Minimum realized acceleration diagnostics.}
\label{fig:min_acceleration_bars}
\vspace{-5mm}
\end{figure}

While Fig.~\ref{fig:efficiency_trends_summary} focuses on efficiency, Figs.~\ref{fig:hard_braking_bars}--\ref{fig:min_acceleration_bars} examine the executed-control quality. Fig.~\ref{fig:hard_braking_bars} reports the frequency of realized acceleration below $-6\,\mathrm{m/s^2}$, and Fig.~\ref{fig:min_acceleration_bars} reports the most severe realized acceleration. These diagnostics are consistent with Table~\ref{tab:main_summary}: DRIFT keeps zero hard-braking events in \textit{Ring}, reduces hard braking relative to Flow-AIL in \textit{F8}, and sharply lowers hard braking relative to the high-throughput baselines in \textit{Merge}. The minimum-acceleration results show the same interpretation from the severity side: DRIFT avoids relying on very large negative acceleration to obtain its efficiency values, so its generated candidates remain more compatible with executable closed-loop rollout.

\subsubsection{Module and Mechanism Analysis}\label{subsec:ablation_analysis}

\begin{figure}[!t]
	% \vspace{-5mm}
	\centering
	\includegraphics[width=\columnwidth]{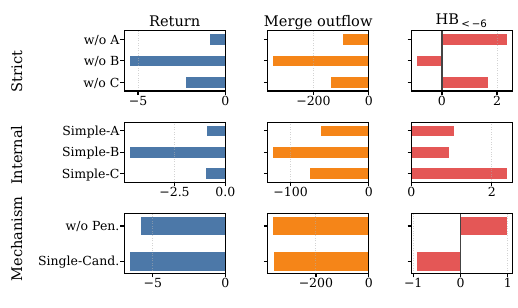}
	\vspace{-10mm}
	\caption{Performance deltas relative to full DRIFT.}
	\label{fig:ablation_delta_summary}
	\vspace{-5mm}
\end{figure}

Fig.~\ref{fig:ablation_delta_summary} examines which design choices are responsible for the aggregate behavior observed above. The `strict' ablations remove Modules A, B, and C, respectively. Removing Module A weakens the heterogeneity-aware conditioning, removing Module B replaces the learned executable diffusion generator, and removing Module C removes discriminator-based behavior similarity and long-tail feedback; in all three cases, at least one of return, speed, outflow, HB, or Worst $a$ moves away from the full DRIFT operating point. The `internal' replacements show that plain conditioning, prior-sampled candidates, and heuristic selection do not reproduce the full learned pipeline. Finally, the `mechanism' ablations isolate penetration conditioning and candidate multiplicity: removing penetration conditioning or restricting the generator to a single candidate substantially lowers return, speed, and \textit{Merge} outflow, even when some braking-related indicators become more conservative. Thus, the ablation results support the role of heterogeneity-aware representation, executable multi-candidate generation, and risk-aware online selection as a coupled design rather than as interchangeable components.

Overall, the experimental results support the central design philosophy of DRIFT. Across the considered scenarios and AV penetration levels, the framework preserves competitive efficiency while reducing several indicators of unsafe or difficult execution. The ablation studies further indicate that these outcomes arise from the interaction among heterogeneity-aware representation learning, executable diffusion-based candidate generation, Module-C feedback, and risk-aware online selection, rather than from a single isolated component.

% Fig.~\ref{fig:ablation_delta_summary} reports changes in the metrics relative to full DRIFT pipeline for strict module ablations (`Strict'), internal replacements (`Internal'), and mechanism ablations (`Mechanism'). 
% % The detailed tables are in Appendix~\ref{sec:appendix_supp_results}. 
% % 
% % The bars in Fig.~\ref{fig:ablation_delta_summary} should be read as metric changes after removing or replacing one part of the full pipeline. 
% The `Strict' group tests the three modules defined in Sections~\ref{sec:module_a}--\ref{sec:module_c}: removing Module A degrades the heterogeneity-aware conditioning, removing Module B removes the learned multi-candidate generator, and removing Module C removes discriminator-based behavior similarity and long-tail reweighting. The `Internal' group compares full DRIFT with the Simple-A/B/C replacements defined in Section~\ref{subsec:baselines_variants}, showing whether plain conditioning, prior-sampled candidates, or heuristic selection can substitute for the learned modules. The `Mechanism' group isolates penetration conditioning, candidate multiplicity, and the online selector $\mathcal{P}_{\mathsf{sel}}$ defined in~\eqref{eq:psel}. Across these groups, the return/speed/outflow bars quantify efficiency changes, while HB and Worst $a$ quantify safety/executability changes. Appendix~\ref{sec:appendix_supp_results} provides the corresponding numerical tables, candidate-set-size sensitivity, latency, selector/feasibility variants, stress tests, Module-C risk-weight sensitivity, and validation-set candidate-realism diagnostics.

% Taken together, the experiments indicate that DRIFT balances safety and efficiency: external baselines can be competitive on isolated efficiency metrics, but Table~\ref{tab:main_summary} and Figs.~\ref{fig:hard_braking_bars}--\ref{fig:min_acceleration_bars} show that they may produce more hard-braking events or more severe realized deceleration. The module and mechanism ablations further show that Modules A/B/C each affect the final closed-loop behavior.}

\section{Conclusion and Future Work}\label{sec:conclusion}
This paper developed DRIFT, a closed-loop traffic-generation framework for mixed-autonomy settings in which HVs and AVs coexist under varying AV penetration rates. DRIFT addresses three coupled challenges: representing heterogeneous traffic states as the traffic composition changes, generating behavior candidates that remain executable under vehicle dynamics and roadway constraints, and incorporating rare safety-critical events into traffic generation and evaluation. To this end, DRIFT combines heterogeneity-aware condition encoding, conditional diffusion-based generation of candidate control sequences, rollout-based feasibility filtering, progressive adversarial alignment with expert driving trajectories, and long-tail safety reweighting.
The experimental evaluation demonstrated the value of this closed-loop design compared with various baselines and internal variants. 
% The present validation is limited to stylized \textit{Flow}/\textit{SUMO} benchmark scenarios, and the AV behavior prior is represented through controlled synthetic scaling rather than real AV-fleet logs. 
Various promising directions follow from this work, such as (i) extending the current vehicle-centered representation and candidate-generation process to larger road networks with more vehicles and more complex interaction structures, and (ii) improving the fidelity of the rollout model and the long-tail feedback mechanism using richer traffic data to better capture complex human-driving responses and rare interaction patterns.

\bibliographystyle{IEEEtran}
\bibliography{references}

\clearpage
\appendices
\section{Additional Existing Literature Positioning}\label{sec:appendix_related_work}
This appendix complements Section~\ref{sec:core_motivation} by providing a compact discussion of representative related works and a scope comparison against DRIFT. The goal is not to exhaustively survey all traffic-generation studies, but to clarify how the main streams of prior work relate to the three challenges considered in this paper: mixed-autonomy heterogeneity, executable closed-loop behavior generation, and risk-aware traffic-level evaluation.

\noindent
$\bullet$~\textit{Imitation Learning and Driving-Behavior Generation:} Mixed-autonomy traffic generation can be viewed as learning executable vehicle-control behaviors that induce traffic-level trajectory evolution. Flow \cite{flow2022} demonstrated the feasibility of learning-based control in mixed-autonomy environments, while PS-TrajGAIL \cite{urbangail2024}, Ctx2TrajGen \cite{ctx2trajgen2025}, multi-agent imitation learning \cite{sunkim2024}, and DiffAIL \cite{diffail2024} improved behavioral distribution matching and interaction modeling. These methods provide useful behavior-learning foundations, but they typically do not jointly model dynamically varying AV penetration, candidate-level executability, and traffic-level closed-loop evaluation.

\noindent
$\bullet$~\textit{Diffusion-Based Traffic and Trajectory Generation:} Diffusion models have been used for controllable trajectory generation, roadway-constrained generation, closed-loop planning, and scene-level traffic synthesis. Representative examples include ControlTraj \cite{controltraj2024}, Diff-RNTraj \cite{diffrntraj2024}, Diffusion-Planner \cite{zheng2025fg}, CDPT \cite{chen2026cdpt}, DiffScene \cite{diffscene2024}, SceneControl \cite{scenecontrol2024}, SceneDiffuser \cite{scenediffuser2024}, LD-Scene \cite{peng2026ldscene}, OGD/ECM-guided generation \cite{jointdiff2024}, intention-aware diffusion prediction \cite{intentiondiff2024}, and CATF \cite{contextaware2024}. These studies highlight the diversity and controllability of diffusion-based generation, whereas DRIFT focuses on generating executable control candidates whose induced trajectories are filtered, scored, and deployed within a repeated simulator-feedback loop.

\noindent
$\bullet$~\textit{Mixed-Autonomy and Heterogeneous Traffic Modeling.} Existing work has studied mixed-autonomy traffic from system-level and data-driven perspectives, including cooperative game-theoretic driving \cite{fang2024mixed}, robust AV--HV traffic control \cite{gongzhu2024}, hybrid-system stability \cite{li2024hybrid}, safety-aware mixed-traffic learning and control \cite{wang2024mixedsafe}, trust-dependent capacity analysis \cite{pei2024capacity}, takeover-induced stability \cite{wang2025transfer}, mixed-autonomy flow and energy impacts \cite{cen2025mixedflow,wang2025energy}, heterogeneous platoons \cite{ren2024platoons}, interaction-aware prediction \cite{zhang2025style}, post-interactive prediction \cite{huang2025pioformer}, and transformer-based MARL \cite{li2025stfusion}. DRIFT differs by connecting microscopic heterogeneous behavior representation with executable candidate generation and macroscopic closed-loop traffic evolution.

\noindent
$\bullet$~\textit{Long-Tail Safety and Scenario Generation.} Long-tail safety studies address rare but high-impact events through dynamic test generation \cite{jia2024dynamictest}, edge-case synthesis \cite{jing2025edge}, corner-case detection \cite{schicktanz2025corner}, uncertainty-aware out-of-distribution safety assessment \cite{wang2025ood}, closed-loop search \cite{trafficmcts2025}, controllable traffic generation \cite{dragtraffic2024}, and adversarial scenario synthesis \cite{peng2026ldscene}. These works improve risk discovery and scenario construction; DRIFT instead incorporates risk information into candidate scoring, long-tail feedback, and traffic-level closed-loop evaluation.

\begin{table}[!t]
	\centering
	\caption{Scope positioning of representative related methods.}
	\label{tab:related_positioning}
	\begingroup
	\setlength{\tabcolsep}{2.0pt}
	\renewcommand{\arraystretch}{0.95}
	\footnotesize
	\begin{tabularx}{\columnwidth}{@{}>{\raggedright\arraybackslash}Xccccc@{}}
		\toprule
		Work & CL & HV/AV & Ctrl. & Risk & Flow/SUMO \\
		\midrule
		Flow \cite{flow2022} & $\checkmark$ & $\checkmark$ & $\checkmark$ & $\circ$ & $\checkmark$ \\
		SceneDiffuser \cite{scenediffuser2024} & $\checkmark$ & $\times$ & $\circ$ & $\circ$ & $\times$ \\
		TrafficMCTS \cite{trafficmcts2025} & $\checkmark$ & $\circ$ & $\times$ & $\checkmark$ & $\times$ \\
		DiffAIL \cite{diffail2024} & $\circ$ & $\times$ & $\times$ & $\circ$ & $\times$ \\
		ControlTraj \cite{controltraj2024} & $\times$ & $\times$ & $\circ$ & $\times$ & $\times$ \\
		CDPT \cite{chen2026cdpt} & $\circ$ & $\times$ & $\times$ & $\circ$ & $\times$ \\
		DragTraffic \cite{dragtraffic2024} & $\circ$ & $\times$ & $\circ$ & $\checkmark$ & $\times$ \\
		LD-Scene \cite{peng2026ldscene} & $\circ$ & $\times$ & $\circ$ & $\checkmark$ & $\times$ \\
		DRIFT & $\checkmark$ & $\checkmark$ & $\checkmark$ & $\checkmark$ & $\checkmark$ \\
		\bottomrule
	\end{tabularx}
	\parbox{\columnwidth}{\vspace{1mm}\scriptsize CL: closed-loop rollout/evaluation; HV/AV: explicit mixed-autonomy heterogeneity; Ctrl.: executable control or candidate generation; Risk: risk-aware or long-tail feedback; Flow/SUMO: direct compatibility with the executable \textit{Flow}/\textit{SUMO} interface used in our experiments. Symbols $\checkmark$, $\circ$, and $\times$ indicate direct support, partial coverage, and not a primary focus, respectively.}
	\endgroup
\end{table}

\section{Overall Workflow of DRIFT}\label{sec:appendix_workflow}
% This appendix gives the operational workflow of DRIFT omitted from the compact overview in Section~\ref{sec:design_drift}. 
Algorithm~\ref{alg:drift_overall} summarizes the offline stagewise training procedure and the online closed-loop inference procedure of DRIFT (the stopping criterion used in line~\ref{line:7} is the episode termination, i.e., the simulation horizon is reached or the scenario-specific terminal condition is met).

\begin{algorithm}[h]
	\caption{Overall training and inference workflow of DRIFT}
	\label{alg:drift_overall}
	\footnotesize
	\Input{training time-vehicle samples $\{(t,i,\mathbf{s}_t^i)\}$, scenario library, replanning set $\mathcal{T}^{\mathsf{replan}}$, horizons $H^{\mathsf{plan}}$ and $H^{\mathsf{exec}}$, initial parameters $\phi_A,\phi_B,\vartheta$}
	\Output{trained parameters $\phi_A,\phi_B,\vartheta$, cached Module-C reweighting coefficients, and an online closed-loop decision policy}
	\BlankLine
	\tcp{Offline stagewise training}
	\Repeat{the stopping criterion is met}{
		sample a time-vehicle mini-batch $\{(t,i)\}$ from the training pool\;
		update $\phi_A$ with the Module-A representation objective on the sampled mini-batch\;
		freeze the updated encoder and update $\phi_B$ with the Module-B denoising and feasibility objectives\;
		freeze $\phi_B$ and update $\vartheta$ with the progressively activated alignment objective\;
		compute long-tail coefficients $\chi_t^{i,k}$ from Module C and set the next Module-B weights $w_t^{i,k}\leftarrow\chi_t^{i,k}$\;
	}
	\label{line:7}
	\BlankLine
	\tcp{Online closed-loop inference}
	\ForEach{RI $t\in\mathcal{T}^{\mathsf{replan}}$}{
		construct the current traffic state $\mathbf{X}_t$ and identify the active vehicle set $\tilde{\mathcal{V}}_t$\;
		\ForEach{active vehicle $i\in\tilde{\mathcal{V}}_t$}{
			form the heterogeneous local state $\mathbf{s}_t^i$\;
			encode the condition vector $\mathbf{c}_t^i=g_{\phi_A}(\mathbf{s}_t^i)$\;
			generate $K$ candidates via Module B and roll them out into candidate trajectories\;
			form $\mathcal{K}_{t}^{i,\mathsf{feas}}$ from $\Omega_t^{i,\mathsf{feas}}$ and evaluate $\{J_t^{i,k}\}_{k\in\mathcal{K}_{t}^{i,\mathsf{feas}}}$ when the set is nonempty\;
			select $k^*=\arg\max_{k\in\mathcal{K}_{t}^{i,\mathsf{feas}}}J_t^{i,k}$, or apply the deterministic fallback in Section~\ref{sec:problem_decomposition} if $\mathcal{K}_{t}^{i,\mathsf{feas}}=\emptyset$\;
			execute the first $H^{\mathsf{exec}}$ PSs of the selected candidate\;
		}
		collect environmental feedback and advance the system to the next RI\;
	}
\end{algorithm}

\section{Implementation Details}\label{sec:appendix_implementation}
This appendix records the implementation settings that are independent of the final numerical results. Time-vehicle samples are constructed from the naturalistic trajectory datasets used for representation and generator training, with a history length of $L_h=6$ and a future/control horizon of six discrete steps. The Module-A encoder uses a temporal history branch and a feed-forward context branch; in the implementation used for this manuscript, the feature dimension is $7$, the context dimension is $6$, the hidden dimension is $96$, and the condition embedding dimension is $48$. The coarse grouping task uses $13$ behavioral-interaction tags. The stability and separation terms in the representation objective are implemented with training-script weights $0.1$ and $0.1$, respectively, with margin $m^{\mathsf{sep}}=1.0$, batch size $256$, Adam optimizer, learning rate $10^{-3}$, weight decay $10^{-5}$, and $20$ training epochs.

Module B generates $K=5$ candidates per RI-vehicle query using a compact executable-control parameterization with speed, spacing, and caution-related attributes. The conditional denoising backbone is implemented with feed-forward layers of hidden dimension $160$, batch size $256$, Adam optimizer, learning rate $10^{-3}$, weight decay $10^{-5}$, and $30$ training epochs. The feasibility regularization weight is set to $\lambda^{\mathsf{fea}}=0.1$ in the implementation. Module C uses the same condition embedding dimension, a hidden dimension of $160$, a realism head and a risk head, batch size $256$, Adam optimizer, learning rate $5\times 10^{-4}$, weight decay $10^{-5}$, and $20$ training epochs. During online inference, the learned Module-C score uses realism minus $\lambda^{\mathsf{risk}}$ times risk, with the default and sensitivity settings reported in Table~\ref{tab:scoring_parameters}.

\begin{table}[!t]
	\centering
	\caption{Executable scoring and implementation parameters.}
	\label{tab:scoring_parameters}
	\begingroup
	\setlength{\tabcolsep}{2.0pt}
	\renewcommand{\arraystretch}{0.94}
	\scriptsize
	\begin{tabularx}{\columnwidth}{p{0.33\columnwidth}X}
\toprule
Item & Value / setting \\
\midrule
Number of candidates & $K=5$ \\
Target speed & $v^{\mathsf{des}}=20.0$ m/s \\
Nominal safe headway & $10.0$ m \\
Default TTC guard & $2.0$ s \\
\textit{Merge} TTC guard & $2.8$ s \\
Module-C risk weight & $\lambda^{\mathsf{risk}}=0.65$ \\
Sensitivity values & $\lambda^{\mathsf{risk}}\in\{0.35,0.65,0.95\}$ \\
\midrule
Heuristic score & $0.90e+0.95p+0.60s-w^{\mathsf{risk}}r-w^{\mathsf{ttc}}q^{\mathsf{ttc}}-w^{\mathsf{gap}}q^{\mathsf{gap}}-q^{\mathsf{conf}}-w^{\mathsf{acc}}q^{\mathsf{acc}}$ \\
Default penalties & $(w^{\mathsf{risk}},w^{\mathsf{ttc}},w^{\mathsf{gap}},w^{\mathsf{acc}})=(1.25,1.10,0.90,0.08)$ \\
Full-penetration guard & $(w^{\mathsf{risk}},w^{\mathsf{ttc}},w^{\mathsf{gap}},w^{\mathsf{acc}})=(1.45,1.30,1.05,0.11)$ \\
\textit{Merge} conflict penalty & $q^{\mathsf{conf}}=0.04$ when the conflict-zone guard is active \\
Learned-score weight & $1.10$ in general; $0.65$ in calibrated \textit{Merge}; $0.55$ under the full-penetration guard \\
\midrule
Training seed & $42$ \\
Module-A epochs & $20$ \\
Module-B epochs & $30$ \\
Module-C epochs & $20$ \\
Feasibility regularization & $\lambda^{\mathsf{fea}}=0.1$ \\
\bottomrule
\end{tabularx}

	\endgroup
\end{table}

\begin{table}[!t]
	\centering
	\caption{AV-prior scaling ratios relative to the calibrated HV prior. Here $v$, $T$, $s_0$, $a$, $b$, $\tau$, and $\sigma$ denote desired speed, time headway, minimum gap, maximum acceleration, comfortable deceleration, reaction delay, and control noise.}
	\label{tab:av_prior_scaling}
	\begingroup
	\setlength{\tabcolsep}{1.6pt}
	\renewcommand{\arraystretch}{0.94}
	\scriptsize
	\begin{tabular}{@{}>{\raggedright\arraybackslash}p{0.14\columnwidth}*{7}{>{\centering\arraybackslash}p{0.104\columnwidth}}@{}}
\toprule
Scen. & $v$ & $T$ & $s_0$ & $a$ & $b$ & $\tau$ & $\sigma$ \\
\midrule
Ring & 1.02 & 0.92 & 0.95 & 1.18 & 1.18 & 0.70 & 0.40 \\
F8 & 1.03 & 0.95 & 1.00 & 1.15 & 1.20 & 0.72 & 0.42 \\
Merge & 1.05 & 0.88 & 0.90 & 1.28 & 1.28 & 0.65 & 0.35 \\
\bottomrule
\end{tabular}

	\endgroup
\end{table}

\begin{table}[!t]
	\centering
	\caption{Experimental protocol and metrics.}
	\label{tab:experiment_protocol}
	\begingroup
	\setlength{\tabcolsep}{1.8pt}
	\renewcommand{\arraystretch}{0.96}
	\footnotesize
	\begin{tabular}{@{}>{\raggedright\arraybackslash}p{0.29\columnwidth}>{\raggedright\arraybackslash}p{0.66\columnwidth}@{}}
\toprule
Item & Setting \\
\midrule
Simulation platform & \textit{Flow} + \textit{SUMO} closed-loop simulation \\
Scenarios & \textit{ring}, \textit{F8}, and \textit{merge} \\
AV penetration & $0/20/40/60/80/100\%$ \\
Runs per cell & 5 closed-loop episodes \\
Efficiency metrics & Return $\uparrow$, average speed $\uparrow$, merge outflow $\uparrow$ \\
Safety diagnostics & Collision/deadlock $\downarrow$, hard braking $\downarrow$, min acceleration $\uparrow$, THW/TTC violations $\downarrow$ \\
\bottomrule
\end{tabular}

	\endgroup
\end{table}

\begin{table}[!t]
	\centering
	\caption{Core implementation settings.}
	\label{tab:reproducibility_protocol}
	\begingroup
	\setlength{\tabcolsep}{1.8pt}
	\renewcommand{\arraystretch}{0.96}
	\footnotesize
	\begin{tabular}{@{}>{\raggedright\arraybackslash}p{0.32\columnwidth}>{\raggedright\arraybackslash}p{0.63\columnwidth}@{}}
\toprule
Item & Implementation detail \\
\midrule
Input window & History/future windows of 6 steps with stride 5; online AV state uses 7 local kinematic and safety features. \\
Module A & Temporal history branch and context MLP; embedding dimension 48; trained for 20 epochs. \\
Module B & Conditional denoising candidate generator with feed-forward backbone; $K=5$ executable-control candidates parameterized by speed, spacing, and caution-related attributes; trained with denoising and feasibility losses. \\
Module C & Realism/risk scorer with candidate and fusion MLPs; learned score is realism $-0.65\times$ risk by default, where risk supervision uses THW, TTC, and lane-change indicators. \\
Safety logs & THW violation: THW $<1.0$ s; TTC violation: TTC $<2.0$ s; hard braking: acceleration $<-6/-10/-20$ m/s$^2$. \\
Runtime check & Server profiling gives $0.783$ ms/action p95 for the executable A/B/C stack at $K=5$; end-to-end \textit{Flow}/\textit{SUMO} throughput is $6.16$--$145.19$ steps/s across profiled closed-loop cells. \\
Robustness check & A 10-run support matrix covers FollowerStopper, PI, and DRIFT over Ring/F8/Merge and $0/20/40/60/80/100\%$ penetration; it supports paired Wilcoxon/\textit{t}-test trend checks rather than replacing the adopted 5-run main table. \\
Flow-RL & PPO/RLlib baseline trained for 50 iterations; checkpoint 000050 is used for 5-run closed-loop evaluation; $0\%$ penetration is N/A. \\
\bottomrule
\end{tabular}

	\endgroup
\end{table}

\begin{table}[!t]
	\centering
	\caption{Method categories, coverage, and implementation status.}
	\label{tab:method_status}
	\begingroup
	\setlength{\tabcolsep}{1.8pt}
	\renewcommand{\arraystretch}{0.96}
	\footnotesize
	\begin{tabular}{@{}>{\raggedright\arraybackslash}p{0.28\columnwidth}>{\raggedright\arraybackslash}p{0.24\columnwidth}>{\raggedright\arraybackslash}p{0.17\columnwidth}>{\raggedright\arraybackslash}p{0.22\columnwidth}@{}}
\toprule
Method & Role & Coverage & Status \\
\midrule
FollowerStopper & Classic rule-based & Six penetrations & Flow-native baseline \\
PI with saturation & Classic rule-based & Six penetrations & Flow-native baseline \\
Flow-RL & Official learning baseline & $20$--$100\%$ & $0\%$ is N/A \\
IDM & Classic car-following & Six penetrations & Flow-native baseline \\
Flow-AIL & Adapted imitation baseline & Six penetrations & Supplementary learning baseline \\
DRIFT & Proposed method & Six penetrations & Full A+B+C stack \\
w/o A/B/C & Strict ablation & Six penetrations & Remove one learned module \\
Simple-A/B/C & Internal replacement & Six penetrations & BC+simple A; AC+simple B; AB+simple C \\
w/o Pen., Single-Cand. & Mechanism ablation & Six penetrations & Disable one design mechanism \\
\bottomrule
\end{tabular}

	\endgroup
\end{table}

In Table~\ref{tab:scoring_parameters}, $e$, $p$, $s$, $r$, $q^{\mathsf{ttc}}$, $q^{\mathsf{gap}}$, $q^{\mathsf{conf}}$, and $q^{\mathsf{acc}}$ denote the normalized efficiency, progress, safety-margin, risk, TTC-violation, close-gap, conflict, and smoothness terms, respectively. The score-component weights in Eq.~(\ref{eq:score_j}) and Appendix~\ref{sec:appendix_scores} are paper-level grouping weights used to define the score components; their operational effect is realized by the executable controller coefficients in Table~\ref{tab:scoring_parameters}. The conceptual $\beta^{\mathsf{lt}}$ and $E^{\mathsf{warm}}$ in Section~\ref{sec:module_c} describe the stagewise reweighting and progressive alignment schedule; in the reported executable implementation, their effect is represented by the fixed Module-C training stage and the runtime risk-weight sweep, rather than by an additional independent grid. The A/B/C stages are trained for fixed epoch budgets with validation-loss monitoring, and the best available checkpoint is used for closed-loop evaluation. All random shuffling of training shards uses a fixed seed of $42$ unless otherwise stated. The external naturalistic trajectory datasets used for calibration are publicly available from their original providers. To support reproducibility, the processed experiment scripts, configurations, and lightweight result tables can be made available upon publication or reasonable request, subject to dataset-provider terms and project release constraints.

\section{Supplementary Results}\label{sec:appendix_supp_results}
This appendix reports auxiliary diagnostics referenced in Section~\ref{subsec:metrics}.

\begin{table}[!t]
	\centering
	\caption{Statistical support for the main results.}
	\label{tab:main_statistics_support}
	\begingroup
	\setlength{\tabcolsep}{1.6pt}
	\renewcommand{\arraystretch}{0.94}
	\scriptsize
	\begin{tabular}{@{}>{\raggedright\arraybackslash}p{0.10\columnwidth}>{\raggedright\arraybackslash}p{0.17\columnwidth}*{4}{>{\centering\arraybackslash}p{0.165\columnwidth}}@{}}
\toprule
Scen. & Method & Ret. $\uparrow$ & Spd. $\uparrow$ & Out. $\uparrow$ & HB/run $\downarrow$ \\
\midrule
Ring & FS & 18.32 $\pm$ 0.22 & 1.59 $\pm$ 0.02 & -- & 17.4 $\pm$ 1.6 \\
Ring & PI & 11.81 $\pm$ 0.00 & 1.01 $\pm$ 0.00 & -- & 0.0 $\pm$ 0.0 \\
Ring & Flow-RL & 11.59 $\pm$ 0.41 & 1.56 $\pm$ 0.06 & -- & 0.0 $\pm$ 0.0 \\
Ring & DRIFT & 25.41 $\pm$ 0.02 & 2.17 $\pm$ 0.00 & -- & 0.0 $\pm$ 0.0 \\
F8 & FS & 52.64 $\pm$ 1.97 & 4.72 $\pm$ 0.19 & -- & 10.3 $\pm$ 1.9 \\
F8 & PI & 40.50 $\pm$ 0.03 & 3.60 $\pm$ 0.00 & -- & 3.3 $\pm$ 0.2 \\
F8 & Flow-RL & 21.43 $\pm$ 0.68 & 2.87 $\pm$ 0.08 & -- & 1.5 $\pm$ 1.1 \\
F8 & DRIFT & 58.38 $\pm$ 0.39 & 5.17 $\pm$ 0.02 & -- & 4.4 $\pm$ 1.2 \\
Merge & FS & 66.73 $\pm$ 3.49 & 14.31 $\pm$ 0.69 & 1378 $\pm$ 84.9 & 23.4 $\pm$ 13.8 \\
Merge & PI & 42.85 $\pm$ 0.79 & 9.14 $\pm$ 0.15 & 726.0 $\pm$ 26.2 & 32.4 $\pm$ 1.6 \\
Merge & Flow-RL & 52.93 $\pm$ 4.31 & 10.77 $\pm$ 0.89 & 812.4 $\pm$ 69.9 & 3.7 $\pm$ 3.2 \\
Merge & DRIFT & 66.80 $\pm$ 3.45 & 14.49 $\pm$ 0.55 & 1377 $\pm$ 63.9 & 5.3 $\pm$ 2.5 \\
\bottomrule
\end{tabular}

	\endgroup
\end{table}

Table~\ref{tab:ablation_details} reports the detailed module and mechanism values.

\begin{table}[!t]
	\centering
	\caption{Detailed ablations and replacement controls.}
	\label{tab:ablation_details}
	\begingroup
	\setlength{\tabcolsep}{1.0pt}
	\renewcommand{\arraystretch}{0.96}
	\scriptsize
	\begin{tabular}{@{}>{\raggedright\arraybackslash}p{0.17\columnwidth}>{\raggedright\arraybackslash}p{0.23\columnwidth}*{5}{>{\centering\arraybackslash}p{0.102\columnwidth}}@{}}
\toprule
Group & Variant & Ret. $\uparrow$ & Spd. $\uparrow$ & Out. $\uparrow$ & Min $a$ $\uparrow$ & HB $\downarrow$ \\
\midrule
Full & DRIFT & \textbf{50.20} & \textbf{7.28} & \textbf{1377.0} & -97.4 & 16.2 \\
\midrule
Strict & w/o A & 49.30 & 7.08 & 1280.0 & -100.3 & 18.6 \\
Strict & w/o B & 44.68 & 6.05 & 1029.0 & -95.4 & 15.2 \\
Strict & w/o C & 47.88 & 6.93 & 1239.0 & -100.8 & 17.8 \\
\midrule
Replace & Simple-A & 49.24 & 7.09 & 1315.0 & -100.2 & 17.2 \\
Replace & Simple-B & 45.46 & 6.85 & 1254.0 & -101.9 & 17.1 \\
Replace & Simple-C & 49.21 & 7.07 & 1301.0 & -100.2 & 18.6 \\
\midrule
Mechanism & w/o Pen. & 44.35 & 6.04 & 1013.0 & -96.3 & 17.2 \\
Mechanism & Single-Cand. & 43.56 & 5.96 & 1015.0 & -96.4 & 15.2 \\
\bottomrule
\end{tabular}

	\endgroup
\end{table}

Table~\ref{tab:k_sensitivity} shows that candidate diversity helps in \textit{F8} but is not monotonic in \textit{Merge}. At high demand, larger $K$ can degrade TTC and outflow, so $K=5$ is used as a practical default. Table~\ref{tab:runtime_module_latency} reports the corresponding online latency.

\begin{table}[!t]
	\centering
	\caption{Sensitivity to candidate-set size.}
	\label{tab:k_sensitivity}
	\begingroup
	\setlength{\tabcolsep}{1.4pt}
	\renewcommand{\arraystretch}{0.94}
	\scriptsize
	\begin{tabular}{@{}>{\raggedright\arraybackslash}p{0.16\columnwidth}>{\centering\arraybackslash}p{0.13\columnwidth}*{5}{>{\centering\arraybackslash}p{0.122\columnwidth}}@{}}
\toprule
Scen./$p$ & Cand. & Ret. $\uparrow$ & Spd. $\uparrow$ & Out. $\uparrow$ & Min $a$ $\uparrow$ & TTC\% $\downarrow$ \\
\midrule
F8/80 & $K=1$ & 53.7 & 4.61 & -- & -29.7 & \textbf{1.56} \\
F8/80 & $K=3$ & 54.2 & 4.65 & -- & -20.6 & 1.70 \\
F8/80 & $K=5$ & 59.9 & 5.41 & -- & -17.4 & 1.57 \\
F8/80 & $K=10$ & \textbf{60.4} & \textbf{5.47} & -- & \textbf{-16.0} & \textbf{1.48} \\
\addlinespace[1pt]
F8/100 & $K=1$ & 53.5 & 4.55 & -- & -21.0 & 1.50 \\
F8/100 & $K=3$ & 61.0 & 5.56 & -- & \textbf{-11.4} & 1.31 \\
F8/100 & $K=5$ & \textbf{61.1} & \textbf{5.58} & -- & -11.4 & \textbf{1.26} \\
F8/100 & $K=10$ & 60.9 & 5.56 & -- & -13.2 & 1.28 \\
\addlinespace[1pt]
Merge/80 & $K=1$ & 63.9 & 14.04 & 1242 & -81.8 & 2.07 \\
Merge/80 & $K=3$ & 66.9 & 14.75 & \textbf{1422} & -75.8 & 0.06 \\
Merge/80 & $K=5$ & 67.4 & 14.89 & 1404 & -74.8 & 0.07 \\
Merge/80 & $K=10$ & \textbf{70.2} & \textbf{15.31} & 1410 & \textbf{-55.2} & \textbf{0.01} \\
\addlinespace[1pt]
Merge/100 & $K=1$ & 70.5 & \textbf{15.54} & \textbf{1404} & -75.4 & \textbf{0.59} \\
Merge/100 & $K=3$ & \textbf{71.2} & 15.52 & 1374 & \textbf{-53.9} & 1.14 \\
Merge/100 & $K=5$ & 67.6 & 14.83 & 1290 & -77.9 & 2.50 \\
Merge/100 & $K=10$ & 64.7 & 14.34 & 1188 & -74.8 & 3.31 \\
\bottomrule
\end{tabular}

	\endgroup
\end{table}

\begin{table}[!t]
	\centering
	\caption{Online inference latency in ms/action.}
	\label{tab:runtime_module_latency}
	\begingroup
	\setlength{\tabcolsep}{1.4pt}
	\renewcommand{\arraystretch}{0.94}
	\scriptsize
	\begin{tabular}{lrrrr}
\toprule
Component & Mean (ms)$\downarrow$ & P95 (ms)$\downarrow$ & P99 (ms)$\downarrow$ & Max (ms)$\downarrow$ \\
\midrule
Module A encoding & 0.3855 & 0.4002 & 0.4317 & 23.6892 \\
Module B generation (K=5) & 0.0624 & 0.0707 & 0.0776 & 0.2543 \\
Module C scoring (K=5) & 0.2849 & 0.2567 & 0.2929 & 78.6622 \\
A/B/C total (K=5) & 0.7906 & 0.7832 & 0.9220 & 110.6572 \\
Module C scoring (K=1) & 0.1303 & 0.1415 & 0.1499 & 0.2199 \\
Module C scoring (K=3) & 0.2731 & 0.2437 & 0.2676 & 92.5336 \\
Module C scoring (K=10) & 0.2574 & 0.2846 & 0.3055 & 3.0063 \\
\bottomrule
\end{tabular}

	\endgroup
\end{table}

Table~\ref{tab:selector_feasibility_ablation} separates the effects of the online selector and the feasibility filter. In \textit{F8}, single-term selection weakens efficiency, while in \textit{Merge}, removing feasibility can improve short-term throughput at the cost of tail braking. Tables~\ref{tab:ood_stress_profiles} and \ref{tab:ood_stress_summary} report the compact OOD/stress results.

\begin{table}[!t]
	\centering
	\caption{Selector and feasibility deltas at high penetration.}
	\label{tab:selector_feasibility_ablation}
	\begingroup
	\setlength{\tabcolsep}{1.0pt}
	\renewcommand{\arraystretch}{0.94}
	\scriptsize
	\begin{tabular}{@{}>{\raggedright\arraybackslash}p{0.14\columnwidth}>{\raggedright\arraybackslash}p{0.20\columnwidth}*{5}{>{\centering\arraybackslash}p{0.112\columnwidth}}@{}}
\toprule
Scen. & Variant & $\Delta$Ret. & $\Delta$Spd. & $\Delta$Out. & $\Delta$Min $a$ & $\Delta$HB10 \\
\midrule
F8 & Eff. only & -0.18 & 0.00 & -- & -8.59 & 0.0 \\
F8 & No feas. & 0.05 & 0.00 & -- & -4.74 & 0.0 \\
F8 & Realism only & -5.31 & -0.76 & -- & -7.69 & 0.0 \\
F8 & Risk only & -5.15 & -0.75 & -- & -1.29 & 0.0 \\
\addlinespace[1pt]
Merge & Eff. only & 2.44 & 0.46 & 75.0 & -5.63 & 0.5 \\
Merge & No feas. & 0.87 & -0.19 & 30.0 & -28.26 & 13.5 \\
Merge & Realism only & 1.57 & 0.18 & -6.0 & -3.33 & 2.5 \\
Merge & Risk only & 1.33 & 0.05 & -33.0 & 2.09 & 0.5 \\
\bottomrule
\end{tabular}

	\endgroup
\end{table}

\begin{table}[!t]
	\centering
	\caption{OOD/stress perturbation settings.}
	\label{tab:ood_stress_profiles}
	\begingroup
	\setlength{\tabcolsep}{1.0pt}
	\renewcommand{\arraystretch}{0.94}
	\scriptsize
	\begin{tabular}{@{}>{\raggedright\arraybackslash}p{0.20\columnwidth}>{\raggedright\arraybackslash}p{0.73\columnwidth}@{}}
\toprule
Profile & Perturbation setting \\
\midrule
Aggressive HV & $v^{\mathsf{des}}\times1.10$, $T\times0.82$, $s_0\times0.85$, $a^{\mathsf{max}}\times1.18$, $b\times1.05$, $\tau\times1.10$, $\sigma\times1.25$. \\
Reaction delay & $\tau\times1.55$, $\sigma\times1.15$; other HV prior terms remain at the calibrated setting. \\
Merge demand & $v^{\mathsf{des}}\times1.04$, $T\times0.92$, $\sigma\times1.15$; highway inflow $2400$ veh/h and ramp inflow $180$ veh/h. \\
\bottomrule
\end{tabular}

	\endgroup
\end{table}

\begin{table}[!t]
	\centering
	\caption{Compact OOD/stress robustness summary.}
	\label{tab:ood_stress_summary}
	\begingroup
	\setlength{\tabcolsep}{1.0pt}
	\renewcommand{\arraystretch}{0.94}
	\scriptsize
	\begin{tabular}{@{}>{\raggedright\arraybackslash}p{0.16\columnwidth}*{8}{>{\centering\arraybackslash}p{0.095\columnwidth}}@{}}
\toprule
Method & R-rk $\downarrow$ & S-rk $\downarrow$ & Ret. & Spd. & TTC\% $\downarrow$ & HB10 $\downarrow$ & Min $a$ $\uparrow$ & H-rk $\downarrow$ \\
\midrule
Flow-AIL & 1.58 & 1.50 & \textbf{65.6} & \textbf{10.39} & 0.87 & 10.7 & -119.0 & 1.83 \\
IDM & 2.17 & 2.25 & 63.6 & 10.35 & \textbf{0.15} & 9.7 & -114.6 & 1.92 \\
DRIFT & 2.25 & 2.25 & 60.3 & 9.50 & 2.74 & \textbf{9.2} & \textbf{-111.2} & \textbf{1.67} \\
\bottomrule
\end{tabular}

	\endgroup
\end{table}

Table~\ref{tab:risk_weight_sensitivity} shows how the Module-C risk-weight setting changes efficiency and severe-braking diagnostics. In \textit{F8}, w/o C is slightly more efficient but substantially worse in minimum acceleration.

\begin{table}[!t]
	\centering
	\caption{Sensitivity to the Module-C runtime risk weight.}
	\label{tab:risk_weight_sensitivity}
	\begingroup
	\setlength{\tabcolsep}{1.5pt}
	\renewcommand{\arraystretch}{0.94}
	\scriptsize
	\begin{tabular}{@{}>{\raggedright\arraybackslash}p{0.13\columnwidth}>{\raggedright\arraybackslash}p{0.16\columnwidth}*{5}{>{\centering\arraybackslash}p{0.114\columnwidth}}@{}}
\toprule
Scen. & C setting & Ret. $\uparrow$ & Spd. $\uparrow$ & Out. $\uparrow$ & Min $a$ $\uparrow$ & TTC\% $\downarrow$ \\
\midrule
F8 & $\lambda^{\mathsf{risk}}=.35$ & 60.5 & 5.50 & -- & \textbf{-14.8} & 1.53 \\
F8 & $\lambda^{\mathsf{risk}}=.65$ & 60.5 & 5.49 & -- & -14.9 & \textbf{1.44} \\
F8 & $\lambda^{\mathsf{risk}}=.95$ & 60.1 & 5.43 & -- & -15.6 & 1.51 \\
F8 & w/o C & \textbf{60.8} & \textbf{5.54} & -- & -45.0 & 1.61 \\
\addlinespace[1pt]
Merge & $\lambda^{\mathsf{risk}}=.35$ & \textbf{67.2} & \textbf{14.89} & \textbf{1368} & \textbf{-71.8} & \textbf{0.91} \\
Merge & $\lambda^{\mathsf{risk}}=.65$ & 65.6 & 14.54 & 1314 & -74.7 & 1.64 \\
Merge & $\lambda^{\mathsf{risk}}=.95$ & 65.4 & 14.61 & 1335 & -75.2 & 1.14 \\
Merge & w/o C & 66.1 & 14.59 & 1308 & -76.9 & 1.35 \\
\bottomrule
\end{tabular}

	\endgroup
\end{table}

Table~\ref{tab:validation_candidate_diagnostics} evaluates executable-candidate coverage rather than offline ADE/FDE prediction.

\begin{table}[!t]
	\centering
	\caption{Validation-set executable-candidate diagnostics.}
	\label{tab:validation_candidate_diagnostics}
	\begingroup
	\setlength{\tabcolsep}{0.7pt}
	\renewcommand{\arraystretch}{0.94}
	\scriptsize
	\begin{tabular}{@{}>{\raggedright\arraybackslash}p{0.26\columnwidth}>{\centering\arraybackslash}p{0.12\columnwidth}*{4}{>{\centering\arraybackslash}p{0.105\columnwidth}}>{\centering\arraybackslash}p{0.16\columnwidth}@{}}
\toprule
Candidate & Samp. & Fit $\downarrow$ & $\Delta_\mu$ $\downarrow$ & $\Delta_\sigma$ $\downarrow$ & SW $\downarrow$ & Role \\
\midrule
Learned $K=1$ & 2.36M & 1.228 & 0.576 & 0.042 & 0.570 & first cand. \\
Learned $K=3$ & 2.36M & 0.910 & -- & -- & -- & best-of-$K$ \\
Learned $K=5$ & 2.36M & \textbf{0.094} & \textbf{0.041} & \textbf{0.018} & \textbf{0.044} & best-of-$K$ \\
Fixed $K=5$ & 2.36M & 0.595 & 0.270 & 0.030 & 0.306 & template \\
Selected $K=5$ & 2.36M & 0.095 & 0.042 & 0.025 & 0.044 & $\mathcal{P}_{\mathsf{sel}}$ \\
\bottomrule
\end{tabular}

	\endgroup
\end{table}

\section{Key Notations}\label{sec:appendix_notations}
This appendix summarizes the core symbols used in the system model for convenient cross-referencing.

{\scriptsize
	\newcommand{\notationitem}[2]{%
		\noindent
		\parbox[t]{0.31\columnwidth}{\raggedright #1}%
		\hfill
		\parbox[t]{0.64\columnwidth}{\raggedright #2}\par\vspace{2pt}}
	\noindent\makebox[\columnwidth]{\rule{\columnwidth}{\heavyrulewidth}}\par\vspace{2pt}
	\notationitem{\textbf{Symbol}}{\textbf{Meaning}}
	\noindent\makebox[\columnwidth]{\rule{\columnwidth}{\lightrulewidth}}\par\vspace{2pt}
	\notationitem{$t,\mathcal{T},T$}{TS index, TS index set, and final TS index}
	\notationitem{$\mathcal{T}^{\mathsf{replan}}$}{set of RIs}
	\notationitem{$\mathcal{T}^{\mathsf{plan}}_t$}{local planning window defined at the current RI}
	\notationitem{$\mathcal{T}^{\mathsf{exec}}_t$}{local EW defined at the current RI}
	\notationitem{$\Delta t$}{simulation step size}
	\notationitem{$\mathcal{V}=\{1,\dots,i,\dots,|\mathcal{V}|\}$}{set of all vehicles}
	\notationitem{$\tilde{\mathcal{V}}_t,\tilde{\mathcal{V}}_t^{\mathsf{HV}},\tilde{\mathcal{V}}_t^{\mathsf{AV}}$}{in-network vehicle set and its HV/AV partitions}
	\notationitem{$\rho_t$}{AV penetration rate at TS $t$}
	\notationitem{$\eta_t^{i}$}{control-authority marker of vehicle $i$ at TS $t$}
	\notationitem{$\mathbf{x}_t^{i}$}{state vector of vehicle $i$}
	\notationitem{$\mathbf{X}_t$}{joint system state}
	\notationitem{$\mathbf{s}_t^{i}$}{heterogeneous local state of vehicle $i$}
	\notationitem{$\mathbf{h}_t^{i},\mathbf{n}_t^{i},\mathbf{m}_t^{i}$}{ego history, neighbor interaction, and map semantics}
	\notationitem{$L_h$}{history-window length}
	\notationitem{$\mathcal{N}_t^{i}$}{interaction neighborhood of vehicle $i$}
	\notationitem{$\Delta\mathbf{x}_t^{i,j}$}{interaction feature of vehicle $j$ relative to vehicle $i$}
	\notationitem{$\mathrm{THW}_t^{i,j},\mathrm{TTC}_t^{i,j}$}{interaction headway and collision-time features}
	\notationitem{$\mathrm{THW}_{t}^{i,k,\mathsf{min}}$}{minimum THW of candidate $k$ over the PW}
	\notationitem{$\mathrm{TTC}_{t}^{i,k,\mathsf{min}}$}{minimum TTC of candidate $k$ over the PW}
	\notationitem{$d_t^{i,k,\mathsf{min}},d^{\mathsf{safe}}$}{minimum predicted longitudinal/conflict-zone clearance and safety clearance}
	\notationitem{$r_t^{i,k,\mathsf{map}},r_t^{i,k,\mathsf{col}}$}{lane-topology violation rate and predicted collision/conflict penalty}
	\notationitem{$\mathcal{G}^{\mathsf{map}}$}{lane-topology graph with connectivity and conflict edges}
	\notationitem{$\Omega(\mathcal{G}^{\mathsf{map}})$}{trajectory feasible set induced by $\mathcal{G}^{\mathsf{map}}$}
	\notationitem{$\Omega_t^{i,\mathsf{feas}}$}{local feasible CS used in the candidate-selection subproblem}
	\notationitem{$\mathcal{P}_0,\mathcal{P}_1,\mathcal{P}_2,\mathcal{P}_3$}{unified problem and its three subproblems}
	\notationitem{$\mathcal{G}_t^{\mathsf{int}}$}{vehicle interaction graph}
	\notationitem{$\mu(\cdot)$}{mapping that extracts local semantic context from topology}
	\notationitem{$H^{\mathsf{plan}},H^{\mathsf{exec}},K$}{planning-window length, EW length, and candidate number}
	\notationitem{$\mathbf{U}_{t}^{i,k}$}{the $k$-th candidate executable control sequence of vehicle $i$}
	\notationitem{$\boldsymbol{\tau}_{t}^{i,k}$}{trajectory induced by rolling out $\mathbf{U}_{t}^{i,k}$}
	\notationitem{$\mathbf{U}_{t}^{i,k^{*}},\boldsymbol{\tau}_{t}^{i,k^{*}}$}{selected control sequence and its induced trajectory}
	\notationitem{$\mathbf{u}_{t+h-1}^{i,k}$}{executable control input at the $h$-th PS of candidate $k$}
	\notationitem{$\mathbf{u}^{\mathsf{min}},\mathbf{u}^{\mathsf{max}},v^{\mathsf{min}},v^{\mathsf{max}}$}{control-input and speed bounds}
	\notationitem{$\mathbf{u}_t^{i}$}{actual executed control of vehicle $i$}
	\notationitem{$H^{\mathsf{exec}}$}{EW length (in PSs) of each replanning period}
	\notationitem{$\mathbf{U}_t^{\mathsf{exec}},\mathbf{U}_{t}^{\mathsf{exec},*}$}{executed control set and selected executed-control set}
	\notationitem{$f(\cdot)$}{vehicle-level one-step transition mapping}
	\notationitem{$\mathcal{F}(\cdot)$}{closed-loop environment state-transition mapping}
	\notationitem{$J_t^{i,k}$}{integrated score of candidate $k$}
	\notationitem{$S_t^{i,k},E_t^{i,k}$}{behavior-similarity term and efficiency term}
	\notationitem{$R_t^{i,k},D_t^{i,k}$}{risk term and violation-cost term}
	\notationitem{$\omega^{\mathsf{sim}},\omega^{\mathsf{eff}},\omega^{\mathsf{risk}},\omega^{\mathsf{dyn}}$}{weights of the score terms}
	\notationitem{$\Delta^{\mathsf{real}},\Delta^{\mathsf{tail}}$}{rollout-level realism and tail-risk target penalties}
	\notationitem{$\beta^{\mathsf{real}},\beta^{\mathsf{tail}}$}{weights of the realism and tail-risk target penalties}
	\notationitem{$k_t^{i,*}$}{candidate index selected for execution for vehicle $i$ at RI $t$}
	\notationitem{$k^*$}{local shorthand for $k_t^{i,*}$ in a fixed RI-vehicle pair}
	\notationitem{$\mathbf{k}$}{rollout-level sequence of selected execution indices}
	\notationitem{$\gamma$}{long-term reward discount factor}
	\notationitem{$\Theta$}{trainable parameter-block collection $\{\phi_A,\phi_B,\vartheta\}$}
	\notationitem{$\phi_A,\phi_B,\vartheta,\phi_C$}{Module-A parameters, Module-B parameters, Module-C discriminator parameters, and the Module-C block shorthand that also maintains warm-up and cached reweighting quantities}
	\notationitem{$b_t^i$}{coarse behavioral tag used for positive/negative grouping in Module A}
	\notationitem{$\mathbf{z}_{h,t}^{i},\mathbf{z}_{n,t}^{i},\mathbf{z}_{m,t}^{i},\mathbf{z}_{r,t}^{i}$}{branch-specific features in Module A}
	\notationitem{$\varphi_h(\cdot),\varphi_n(\cdot),\varphi_m(\cdot)$}{branch encoders for history, interaction, and map-semantics streams}
	\notationitem{$\mathbf{e}_{\eta}(\cdot),\mathbf{e}_{\rho}(\cdot),\alpha_t^{i,j}$}{heterogeneous embeddings and interaction-attention weight}
	\notationitem{$\mathbf{c}_t^{i}$}{condition vector from Module A}
	\notationitem{$\mathcal{L}^{\mathsf{repr}},\mathcal{L}^{\mathsf{stb}},\mathcal{L}^{\mathsf{sep}}$}{representation-learning objective, intra-class compactness term, and inter-class separation term}
	\notationitem{$\mathbf{y}_0^{i,k},\mathbf{y}_{d}^{i,k}$}{vectorized candidate control sequence and its $d$th noisy state}
	\notationitem{$\widehat{\mathbf{U}}_{t}^{i,k},\widehat{\boldsymbol{\tau}}_{t}^{i,k}$}{reconstructed control sequence and induced trajectory of candidate $k$}
	\notationitem{$L^{\mathsf{diff}},\beta_{d},\alpha_{d},\bar{\alpha}_{d}$}{diffusion steps, noise schedule, signal-preservation coefficient, and accumulated preservation coefficient}
	\notationitem{$\boldsymbol{\xi}^{i,k},\operatorname{Pred}_{\phi_B}(\cdot)$}{Gaussian noise and Module B noise predictor}
	\notationitem{$\mathcal{L}^{\mathsf{gen}},\mathcal{L}^{\mathsf{noise}},\mathcal{L}^{\mathsf{fea}},\lambda^{\mathsf{fea}}$}{generator objective, denoising loss, feasibility loss, and the feasibility weight}
	\notationitem{$w_t^{i,k},\lambda^{\mathsf{safe}},\lambda^{\mathsf{col}}$}{candidate training weight, safety-margin weight, and conflict penalty weight}
	\notationitem{$\lambda^{\mathsf{sep}},\lambda^{\mathsf{reg}},m^{\mathsf{sep}}$}{separability weight, regularization weight, and margin}
	\notationitem{$\varepsilon$}{small constant used to avoid division by zero in THW/TTC computation}
	\notationitem{$\mathcal{S}_t^{i,+},\mathcal{S}_t^{i,-}$}{positive/negative sets of time-vehicle sample $(t,i)$}
	\notationitem{$\lambda^{\mathsf{topo}},\Delta^{\mathsf{topo}}(\cdot),\Omega_{t}^{\mathsf{map}}$}{topological-feasibility weight, PW-level topology-deviation rate, and map-admissible state set}
	\notationitem{$g_{\phi_A}(\cdot),\operatorname{Pred}_{\phi_B}(\cdot)$}{condition encoder and conditional denoising predictor}
	\notationitem{$\boldsymbol{\tau}_{t}^{i,\mathsf{exp}}$}{expert reference trajectory}
	\notationitem{$\mathcal{D}_{\vartheta}(\cdot),\vartheta,\sigma(\cdot)$}{Module C discriminator, discriminator-parameter block, and Sigmoid mapping}
	\notationitem{$\chi_t^{i,k},\beta^{\mathsf{lt}}$}{long-tail reweighting coefficient and gain}
	\notationitem{$\lambda^{\mathsf{adv},(e)},e,E^{\mathsf{warm}}$}{adversarial warm-up weight, epoch, and warm-up length}
	\notationitem{$\mathcal{L}^{\mathsf{align},(e)},\mathcal{L}^{\mathsf{align}},\mathcal{L}^{\mathsf{tail}}$}{progressive alignment loss, base alignment loss, and long-tail weighted discrepancy}
	\noindent\makebox[\columnwidth]{\rule{\columnwidth}{\heavyrulewidth}}\par
}

\section{Detailed Definitions of Closed-Loop Scoring Terms}\label{sec:appendix_scores}
This appendix gives one normalized implementation of the score terms in Eq.~(\ref{eq:score_j}) using the candidate-level quantities defined in the main text. For ease of weighted aggregation, all terms are normalized or clipped to comparable ranges before aggregation. Here, $k\in\{1,\dots,K\}$ denotes the candidate index, and $k^{*}$ denotes the execution index obtained by score maximization. The terms below specify the component-level aggregation, while Appendix~\ref{sec:appendix_implementation} reports the executable coefficients used in the experiments. Thus, the $\alpha^{\mathsf{label}}$ symbols in this appendix should be interpreted as score-component grouping weights, and their operational effect is realized through the normalized online selector coefficients reported in Appendix~\ref{sec:appendix_implementation}.

\subsection{Behavior Similarity Term}
The behavior-similarity term is defined as:
\begin{equation}
	S_t^{i,k}=\sigma\!\left(
	\mathcal{D}_{\vartheta}\!\left(\boldsymbol{\tau}_{t}^{i,k},\mathbf{s}_t^{i}\right)
	\right),
\end{equation}
where $\mathcal{D}_{\vartheta}(\cdot)$ denotes the discriminator scoring function and $\sigma(\cdot)$ is the Sigmoid mapping.

\subsection{Efficiency Term}
The efficiency term is defined as:
\begin{equation}
	\begin{aligned}
		E_t^{i,k}={}&\alpha^{\mathsf{v}}\frac{\bar{v}_t^{i,k}}{v^{\mathsf{ref}}}\\
		&-\alpha^{\mathsf{wait}} r_t^{i,k,\mathsf{wait}},
	\end{aligned}
\end{equation}
where:
\begin{equation}
	\begin{aligned}
		\bar{v}_t^{i,k}&=\frac{1}{H^{\mathsf{plan}}}\sum_{h=1}^{H^{\mathsf{plan}}}v_{t+h}^{i,k},\\
		r_t^{i,k,\mathsf{wait}}&=\frac{1}{H^{\mathsf{plan}}}\sum_{h=1}^{H^{\mathsf{plan}}}\mathbb{I}\!\left(v_{t+h}^{i,k}<v^{\mathsf{th}}\right).
	\end{aligned}
\end{equation}
In these expressions, $\alpha^{\mathsf{v}},\alpha^{\mathsf{wait}}\ge 0$ are weighting coefficients for the average-speed reward and waiting-time penalty, respectively; $\bar{v}_t^{i,k}$ is the mean speed of candidate $k$ over the prediction horizon; $v^{\mathsf{ref}}>0$ is a reference speed used for normalization; $r_t^{i,k,\mathsf{wait}}$ is the low-speed occupancy ratio; and $v^{\mathsf{th}}$ is the waiting-speed threshold.

\subsection{Safety Risk Term}
The safety-risk term is defined as:
\begin{equation}
	\begin{aligned}
		R_t^{i,k}={}&\alpha^{\mathsf{thw}} r_t^{i,k,\mathsf{thw}}\\
		&+\alpha^{\mathsf{ttc}} r_t^{i,k,\mathsf{ttc}},
	\end{aligned}
\end{equation}
where the risk thresholds are defined consistently with Section~\ref{sec:candidate_selection}. Accordingly, the THW- and TTC-based penalties use the same thresholds $\delta^{\mathsf{THW}}$ and $\delta^{\mathsf{TTC}}$ as those in the local feasible candidate set $\Omega_t^{i,\mathsf{feas}}$:
\begin{equation}
	r_t^{i,k,\mathsf{thw}}=\max\!\left(
	0,\frac{\delta^{\mathsf{THW}}-\mathrm{THW}_{t}^{i,k,\mathsf{min}}}{\delta^{\mathsf{THW}}}
	\right),
\end{equation}
\begin{equation}
	r_t^{i,k,\mathsf{ttc}}=\max\!\left(
	0,\frac{\delta^{\mathsf{TTC}}-\mathrm{TTC}_{t}^{i,k,\mathsf{min}}}{\delta^{\mathsf{TTC}}}
	\right).
\end{equation}
Here, $\mathrm{THW}_{t}^{i,k,\mathsf{min}}$ and $\mathrm{TTC}_{t}^{i,k,\mathsf{min}}$ are the candidate-level minimum safety indicators defined in Section~\ref{sec:candidate_selection}. The coefficients $\alpha^{\mathsf{thw}},\alpha^{\mathsf{ttc}}\ge 0$ weight the THW- and TTC-based risk contributions, respectively. The two normalized penalties $r_t^{i,k,\mathsf{thw}}$ and $r_t^{i,k,\mathsf{ttc}}$ become active only when the candidate violates the corresponding safety threshold.

\subsection{Dynamic and Topology Violation Cost}
The dynamic/topological violation cost is defined as:
\begin{equation}
	\begin{aligned}
		D_t^{i,k}={}&\alpha^{\mathsf{acc}} r_t^{i,k,\mathsf{acc}}\\
		&+\alpha^{\mathsf{map}} r_t^{i,k,\mathsf{map}}\\
		&+\alpha^{\mathsf{col}} r_t^{i,k,\mathsf{col}},
	\end{aligned}
\end{equation}
where:
\begin{equation}
	r_t^{i,k,\mathsf{acc}}=\frac{1}{H^{\mathsf{plan}}-1}\sum_{h=2}^{H^{\mathsf{plan}}}
	\left\|\mathbf{u}_{t+h-1}^{i,k}-\mathbf{u}_{t+h-2}^{i,k}\right\|_2^2,
\end{equation}
$r_t^{i,k,\mathsf{map}}$ and $r_t^{i,k,\mathsf{col}}$ are the topology and predicted conflict penalties introduced in Section~\ref{sec:candidate_selection}. In the executable implementation, the predicted conflict penalty uses:
\begin{equation}
	r_t^{i,k,\mathsf{col}}=
	\max\!\left(
	0,\frac{d^{\mathsf{safe}}-d_t^{i,k,\mathsf{min}}}{d^{\mathsf{safe}}}
	\right)
	+\mathbb{I}\!\left(d_t^{i,k,\mathsf{min}}\le 0\right),
\end{equation}
where $d_t^{i,k,\mathsf{min}}$ and $d^{\mathsf{safe}}$ follow the main-text definitions. Thus, the term is activated before an actual simulator collision occurs, using the predicted candidate occupancy and leader/conflict geometry. The weights $\alpha^{\mathsf{acc}},\alpha^{\mathsf{map}},\alpha^{\mathsf{col}}\ge 0$ balance smoothness, topology consistency, and collision avoidance, respectively.

\end{document}